\documentclass[12pt,preprint]{aastex}
\newcommand{\HST}{{\sl HST}}
\newcommand{\picplace}[1]{\vbox{\hrule\@height 0.4pt\@width\hsize
\hbox to\hsize{\vrule\@width 0.4pt\@height#1\hfil
\vrule\@width 0.4pt\@height#1}\hrule\@height 0.4pt\@width\hsize}}

\slugcomment{Published in PASP, {\bf 848}, 1455\,--\,1473 (2006)}

\shorttitle{CTE calibration for the STIS CCD}
\shortauthors{Goudfrooij et al.}

\begin{document}


\title{Empirical Corrections for Charge Transfer Inefficiency  
  and Associated Centroid Shifts for STIS CCD Observations\altaffilmark{1}} 


\author{Paul Goudfrooij, Ralph C. Bohlin, Jes\'us
  Ma\'{\i}z-Apell\'aniz\altaffilmark{2,3}} 
\affil{Space Telescope Science Institute, 3700 San Martin Drive,
  Baltimore, MD 21218} 
\email{goudfroo, bohlin, jmaiz@stsci.edu}

\and

\author{Randy A. Kimble}
\affil{NASA Goddard Space Flight Center, Code 667, Greenbelt, MD 20771}
\email{randy.a.kimble@nasa.gov}


\altaffiltext{1}{Based on observations with the NASA/ESA {\it Hubble
    Space Telescope}, obtained at the Space Telescope Science
    Institute, which is operated by AURA, Inc., under NASA contract
    NAS5-26555.} 
\altaffiltext{2}{Affiliated with the Space Telescope Division,
    European Space Agency.}
\altaffiltext{3}{Current Address: Instituto de Astrof\'{\i}sica de
    Andaluc\'{\i}a, P.\ O.\ Box 3004, E-18080 Granada, Spain}


\begin{abstract}
A variety of on-orbit imaging and spectroscopic observations are 
used to characterize the Charge Transfer Efficiency (CTE) of the
Charge-Coupled Device (CCD) of the Space Telescope Imaging
Spectrograph (STIS) aboard the {\it Hubble Space Telescope}. A set of 
formulae is presented to correct observations of point sources for
CTE-related loss of signal. For data taken in imaging mode, the CTE
loss is parametrized in terms of the location of the source on the
CCD, the source signal level within the measurement aperture, the
background level, and the time of observation. For spectroscopic data, it
is found that one additional parameter is needed to provide an adequate
calibration of the CTE loss, namely the signal in the point spread function
located between the signal extraction box and the read-out amplifier. The
effect of the latter parameter is significant for spectra taken using the
G750L or G750M gratings of STIS.  The algorithms presented here correct
flux calibration inaccuracies due to CTE losses as large as 30\% to within
$\sim$\,1.5\% RMS throughout the wavelength range covered by the STIS CCD
modes. This uncertainty is similar to the Poisson noise associated with a
source detected at a signal level of about 2500 electrons per resolution
element.  Using bi-directional CCD readouts, centroid shifts incurred due
to CTE loss are also derived. A tight correlation is found between the CTE
loss and the centroid shift (both for imaging and spectroscopic modes),
thus enabling one to correct for both effects of imperfect charge
transfer to STIS CCD observations. 
\end{abstract}



%% chosen the first four characters of the first author's name plus
%% "et al" references, and the first three characters of each author
%% plus the last two numerals of the year of publication for
%% two-author papers. 

\section{Introduction}

Astronomical observation was revolutionized about 25 years ago
by charge-coupled device (CCD) technology, due to a combination of
generally linear response over a very large dynamic range and high
quantum efficiency. However, a shortcoming of CCDs is the fact that
the transfer of charge from one pixel to the next is not
perfect. Charge Transfer Efficiency (CTE) is the term commonly used to 
describe this effect, and it is quantified by the fraction of
charge successfully moved (clocked) between adjacent pixels. In
practice it is often more useful to use the term Charge Transfer {\it
Inefficiency\/} (CTI $\equiv$ 1$-$CTE). We will do so in this paper
when considering quantitative measures. The main observational effect
of CTE loss is that a star whose induced charge has to traverse many
pixels before being read out appears to be fainter than the same star
observed near the read-out amplifier.  

Laboratory tests have shown that the CTE loss of CCDs increases significantly when
being subjected to radiation damage \citep[e.g.,][]{jane+91}. This is
particularly relevant for space-borne CCDs such as those aboard {\it Hubble
Space Telescope (HST)}, where the flux of high-energy
particles, particularly in the South Atlantic Anomaly, is significantly
higher than on the ground. The effect is significant for all CCD detectors
used on {\it HST\/} instruments (e.g., WFPC2: Whitmore, Heyer, 
\& Casertano 1999; Dolphin 2000; STIS: Goudfrooij \& Kimble 2003; ACS:
Riess \& Mack 2004), and charge losses increase with elapsed on-orbit time.
The purpose of the current paper is to characterize the
CTE loss of the CCD of the Space Telescope Imaging Spectrograph (STIS) for
observations of point sources in terms of its dependencies on the X and Y
positions, target intensity, background counts, measurement aperture size,
observing mode, and elapsed on-orbit time.   
Several aspects of on-orbit characterizations of the CTE loss of the STIS
CCD have been reported by \citet*{gill+99}, \citet*{kimb+00}, 
\citet*{goukim03}, and \citet*{bohgou03}. The current paper provides a
comprehensive description of the final STIS CTE calibrations, using data
taken through mid-2004 (i.e., just before STIS operations
ended) which allows a more accurate temporal dependence of the CTE loss than
the reports mentioned above. The current paper also includes analysis
of data taken specifically to constrain the CTE loss at lower signal levels than
before. Furthermore, we provide (for the first time) {\it (i)\/}
separate characterizations of the CTE loss for spectroscopic and imaging
modes, {\it (ii)\/} a comparison between data taken in gain = 1
and gain = 4 e$^-$ DN$^{-1}$ settings, and {\it (iii)\/} the relation of
the CTE loss with its associated centroid shifts. 

\begin{figure}
\centerline{%\psfig{file=f1.eps,width=13cm}
\includegraphics[width=0.7\textwidth]{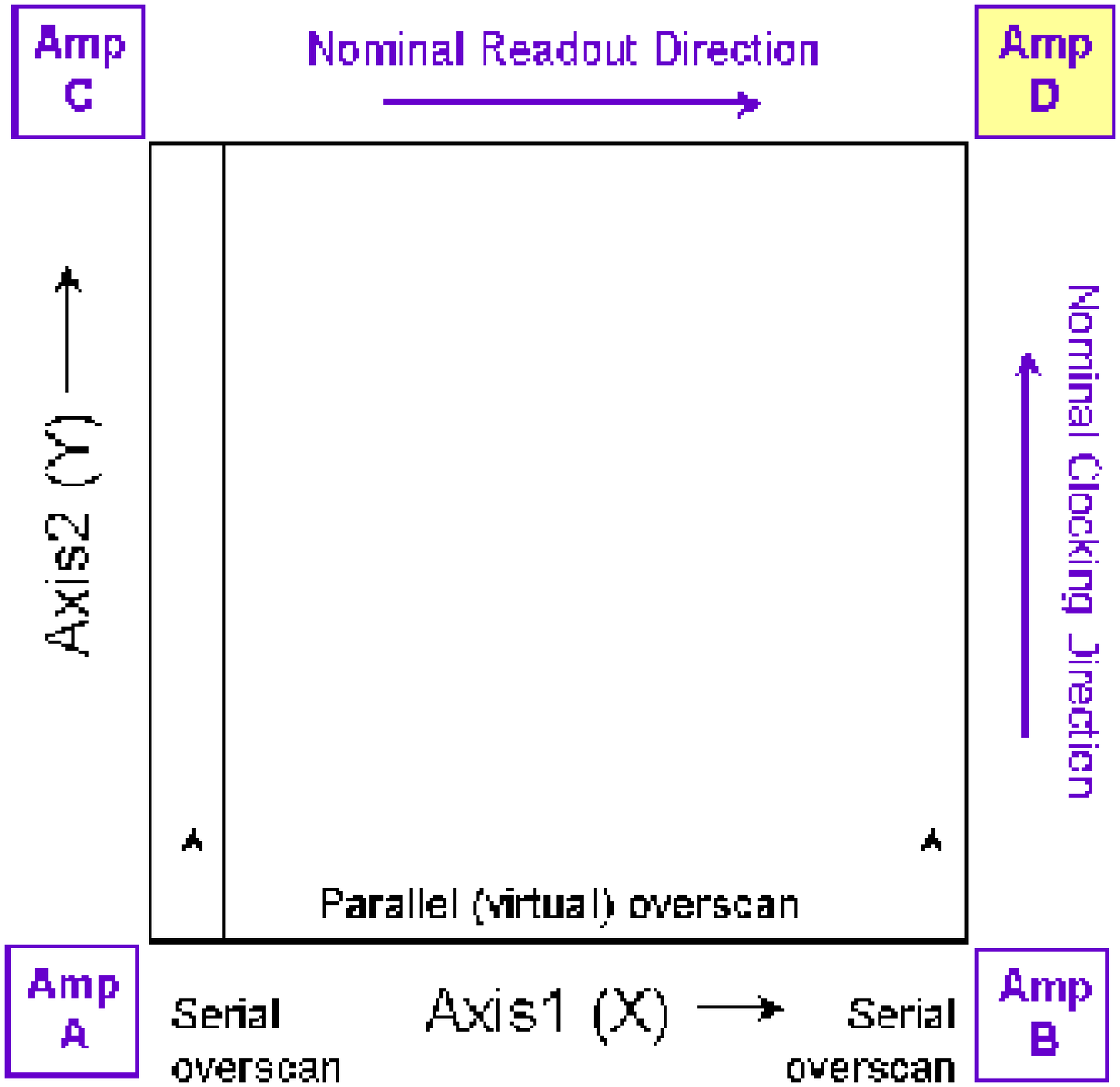}
}
\caption{Schematic architecture of the STIS CCD. The 1024 $\times$
  1024 pixel device has two serial registers and four readout
  amplifiers. The image can be read out through any single amplifier
  or via two-- and four-amplifier combinations. The
  nominal amplifier (amp D) is at the top right.\label{f:ccdarch}}
\end{figure}

The STIS CCD is a 1024 $\times$ 1024 pixel, backside-illuminated device with
21\,$\mu$m $\times$ 21\,$\mu$m pixels. The CCD was built by Scientific Imaging
Technology (SITe) with a coating process that allows coverage of the
200\,--\,1000 nm wavelength range for STIS in a wide variety of imaging and
spectroscopic modes. 
Key features of the STIS CCD architecture are shown schematically in
Figure~\ref{f:ccdarch}. Two serial registers are available. A read-out
amplifier is located at all four corners, each with an independent analog
signal processing chain. The full image can be read out through any one of the
four amplifiers, or through two-- and four-amplifier combinations. By default,
science exposures employ full-frame readout through amplifier `D', which
features the lowest read-out noise (4.0 e$^-$ RMS at launch). 

Further technical details regarding the STIS CCD in particular are provided in
\citet{kimb+94}, while background information on the design of STIS in general
can be found in \citet{wood+98}. 

This paper is organized as follows. Section~\ref{s:moni} describes methods
used to measure the CTE loss:\ Two methods designed to quantify different aspects
of CTE loss for observations of point sources in sparse fields for
spectroscopic and imaging modes, and one method to improve the
characterization of the CTE loss for point source spectroscopy, using on-orbit
observations of spectrophotometric standard stars. Section~\ref{s:results}
presents the results of the CTE analyses, and Section~\ref{s:concl} summarizes
the results. 

\section{CTE Measurement Methods}  \label{s:moni}

\subsection{Spectroscopic Modes}

\subsubsection{Internal Sparse Field Test}    \label{s:ISFtest}

This novel test method, designated the ``internal sparse field'' test,
was developed by the STIS Instrument Definition Team. The method quantifies two key
aspects of CTE effects on spectroscopic measurements:\ {\it (i)\/} The amount
of charge lost {\it outside\/} a standard extraction aperture, and {\it
  (ii)\/} the amount of centroid shift experienced by the charge that remains
{\it within\/} that extraction aperture. 
This test utilizes the ability of the STIS CCD and its associated electronics
to read out the image with any amplifier, i.e., by clocking the accumulated
charge in either direction along both parallel and serial registers. A sequence
of nominally identical exposures is taken, alternating the readout between
amplifiers on opposite sides of the CCD (e.g., amps `B' and `D' for measuring
parallel CTE performance). After correcting for (small) gain differences in
the two readout amplifier chains, the observed ratio of the fluxes 
measured by the two amps can be fit to a simple CTE model of constant
fractional charge loss per pixel transfer. Assigning {\it ybin\/} to be the
CCD binning factor in the parallel clocking (``$y$'') direction and $y$ the
pixel location of the target along the $y$ axis where $y = 1$ at the
first pixel from the location of amp B\footnote{This represents the nominal
  coordinate system of STIS CCD observations in the {\it HST\/} archive}, the
measured ratio of signals measured by amps B and D is   
\begin{equation}
\frac{S_B}{S_D} \, (y) = \frac{\mbox{\it
    CTE}\:^{\mbox{\it y}\,*\mbox{\it ybin}}}{\mbox{\it CTE}\:^{1024 \:- \mbox{\it
    y}\,*\mbox{\it ybin}}} = \mbox{\it CTE}\:^{2 (\mbox{\it y}\,*\mbox{\it ybin}) - 1024}
\label{eq:ctefunc} 
\end{equation}
The dependence of the observed signal ratio (e.g., $S_B/S_D$) on the
source position on the CCD confirms the consistency with a charge transfer
effect (cf.\ Sections~\ref{s:ESFtest} and \ref{s:timedep} below). The CTE is
equal to the exponential of half the slope of ln\,($S_B$/$S_D$) vs.\ $y$. 

A key virtue of this method is that it requires neither a correction for
flat-field response, 
nor an a-priori knowledge of the source flux (as long as the input
source is stable during the alternating 
exposures). It should be noted that what is being measured is a sum
of the charge transfer inefficiencies for the two different clocking
directions. However, for identical clocking voltages and waveforms and with 
the expected symmetry of the radiation damage effects, we believe the
assumption that the CTE loss is equal in the two different clocking
directions is a reasonable one. 

\begin{figure}
\centerline{%\psfig{file=f2.eps,width=16cm,angle=90.}
\includegraphics[width=\textwidth]{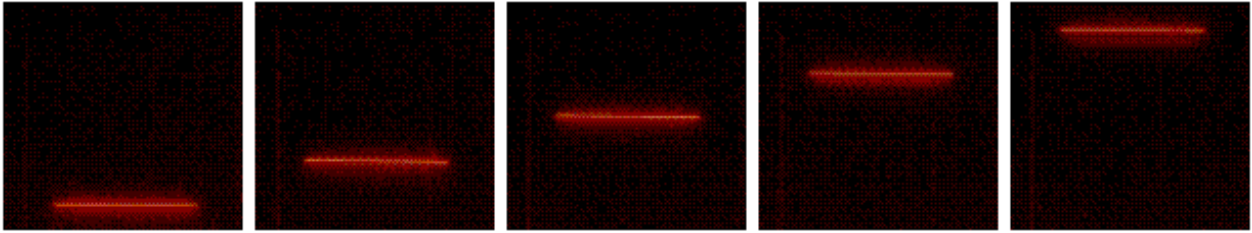}
}
\caption{Representative images used for the ``internal'' sparse field
 CTE test in the parallel clocking direction. At each of the five
 positions along the CCD columns, a sequence of exposures is taken,
 alternating between amplifiers `B' and `D' (see
 Figure~\ref{f:ccdarch}). Systematic variation of the relative signals
 measured by the two amplifiers as a function of position reveals the
 CTE effects. The images are displayed at a logarithmic
 stretch.\label{f:ISFimages}} 
\end{figure}

The implementation of this ``internal''\footnote{``internal'' in this context
  means that all necessary observations use onboard lamps, so that such
  observations can be performed during Earth occultations, hence not requiring
  any valuable ``external'' \HST\ observing time} 
version of the sparse field test is as follows. 
Using an onboard tungsten lamp, the image of a long and narrow slit is
projected at five positions along the CCD columns using special
commanding of the STIS Mode Select Mechanism. At each position, a sequence of
exposures is taken, alternating between the `B' and `D' amplifiers for
readout. An illustration of such an exposure sequence is depicted in
Figure~\ref{f:ISFimages}. The exposure setup used for these observations is
listed in Table~\ref{t:ISFobs}, while the calibration program numbers and
dates of each observing epoch are given in Table~\ref{t:obsblocks}.   

The illumination of these images is representative for typical spectroscopic
observations (as the dispersion direction of STIS CCD spectral modes is
essentially along rows). The slit image has a narrow profile (2-pixel FWHM),
similar to a point source 
spectrum. The CTE loss resulting from this test is ``worst-case'', since there is
hardly any background intensity (``sky'') to provide filling of charge
traps in the CCD array. 

After basic CCD reduction (subtraction of the bias overscan vector 
and bias image subtraction), the average flux per column within a
7-row extraction aperture [which is the default extraction size for
1-D spectral  extractions in the {\sc calstis} pipeline used for STIS
spectra of point sources, cf.\ \citet*{leiboh97,mcgr+99}] as well as the 
centroid of the image profile within those 7 rows are calculated for
each exposure. In order to reflect the flux measurement method used by
the 1-D spectral extraction module of {\sc calstis}, a measurement of
the background level is obtained 40 pixels above and below the
extracted flux with a width of 5 rows. Flux and background 
measurements are performed using $\kappa-\sigma$ clipping in order to
reject cosmic rays and hot pixels. The alternating exposure sequence
allows one to separate CTE effects from flux variations produced by
warmup of the tungsten lamp. As the slit image extends across hundreds
of columns, high statistical precision on CTE performance can 
be obtained even at low signal levels per column. 

We emphasize that in calculating CTE loss from this test, charge is only
considered ``lost'' if it is no longer within the standard 7-row
extraction aperture. I.e., we are only measuring the component of
CTE loss produced by relatively long-time-constant charge trapping. Hence,
the CTI (= 1 $-$ CTE) values derived from this test will not agree with those
measured by (e.g.) X-ray stimulation techniques using Fe$^{55}$ or
Cd$^{109}$, for which charge deferred to even the very first trailing pixel
contributes to the CTI. However, the measurement described here {\it is\/}
directly relevant to the estimation of CTE effects on STIS spectrophotometry.

\subsubsection{Spectrophotometric Sensitivity Monitoring} \label{s:specmoni}

Another set of data we use to characterize the CTE loss of the STIS CCD in
spectroscopic mode is the calibration program that monitors the
sensitivity of STIS spectroscopic modes. This program consists of wide-slit
spectroscopy of primary and secondary spectrophotometric standard stars, using
all supported STIS gratings. These data were taken on a regular basis, with a
monitoring frequency depending on the grating (see \citet*{bohl99} and
\citet*{stys+04} for details). The particular value of this dataset in 
the context of characterizing the CTE loss is that the spectra of these stars
(which are known to be intrinsically constant in time) provide a smoothly
varying signal level along the dispersion direction, thus constraining the
signal-level dependence of the CTE loss particularly well. This is discussed
further in Section~\ref{s:funcforms}.   

\subsection{Imaging Mode}

\subsubsection{External Sparse Field Test} \label{s:ESFtest}

Similar sparse-field CTE tests using ``external'' astronomical data
have also been carried out in flight\footnote{\HST\ Program ID's to
  date are 8415, 8854, 8911, and 9621.}. 
Series of imaging data were acquired on an annual basis for a 
field in the outskirts of the Galactic globular cluster NGC\,6752, a field
containing several hundreds of stars spanning a large range of intrinsic
brightness. Every visit of the field consisted of 3 \HST\ orbits, in which several
exposures were taken using two different exposure times. Several
repeat exposures were taken at both exposure times, alternating again between
opposing readout amplifiers. Details of the imaging observations are listed in
Table~\ref{t:ESFobs}. 

We deliberately chose a sparsely populated region in the outskirts of
NGC\,6752, as it is well known that the CTE-induced loss in crowded
fields is significantly ameliorated (due to trap filling) relative to
the effects on  isolated point sources, while the latter is what we
attempt to measure here.  To allow an assessment of the effect of a
varying sky background level, we 
took the data in the so-called Continuous Viewing Zone (CVZ) of \HST, in
which the bright Earth comes closer than usual to the telescope pointing
direction. The varying amount of scattered light from the bright Earth allows
one to obtain a varying ``sky'' background during the CVZ orbits, and hence to
obtain measurements of CTI at a suitable range of sky background levels. 

Basic CCD reduction steps were performed using tasks in the {\sc
stsdas.hst{\_}calib.stis} package of {\sc iraf}\footnote{{\sc iraf\/} is
 distributed by the National Optical Astronomy Observatories, which is
 operated by the Association of Research in Astronomy, Inc., under
 cooperative agreement with the National Science Foundation, U.S.A.}. After
performing bias and dark subtraction and flatfielding of each individual image
using task {\sc basic2d}, each set of images taken with a given exposure time
is divided up in subgroups according to the sky background value of the
individual images. The sky levels are derived by calculating the centroid of the
histogram of sky pixel values windowed to $\pm 4 \sigma$ of the mean. This is 
done using an iterative procedure which rejects all pixels outside that
window until the number of rejected pixels stays constant from one iteration
to the next. Each subgroup is subsequently summed together (while
performing cosmic-ray rejection; task {\sc ocrreject}), and the summed image
is divided by the number of images combined. The sky background values of the
resulting ``final'' images are listed in Table~\ref{t:ESFobs}.  

Aperture photometry is performed using the {\sc daophot-ii} package
\citep{stet87} as implemented within {\sc iraf} 
using fixed-size apertures. Three aperture sizes are used (radii of 2,
3, and 5 pixels) to evaluate the dependence of calculated CTI values on
aperture size (see Section~\ref{s:results}). An initial star coordinate
list was established using the {\sc daofind} task, employing a detection
threshold of 5$\sigma$ above the background. This initial list was then
cleaned by eliminating targets close to bright stars, in order to avoid
anomalous CTI measurements due to excessive filling of charge 
traps. Saturated stars were also eliminated. This cleaned coordinate list
was used for all measurement epochs, using {\sc iraf} task {\sc xyxymatch} to
calculate the appropriate shifts and rotation angles to be applied to the
reference coordinate list for each epoch. 

\begin{figure}
\plotone{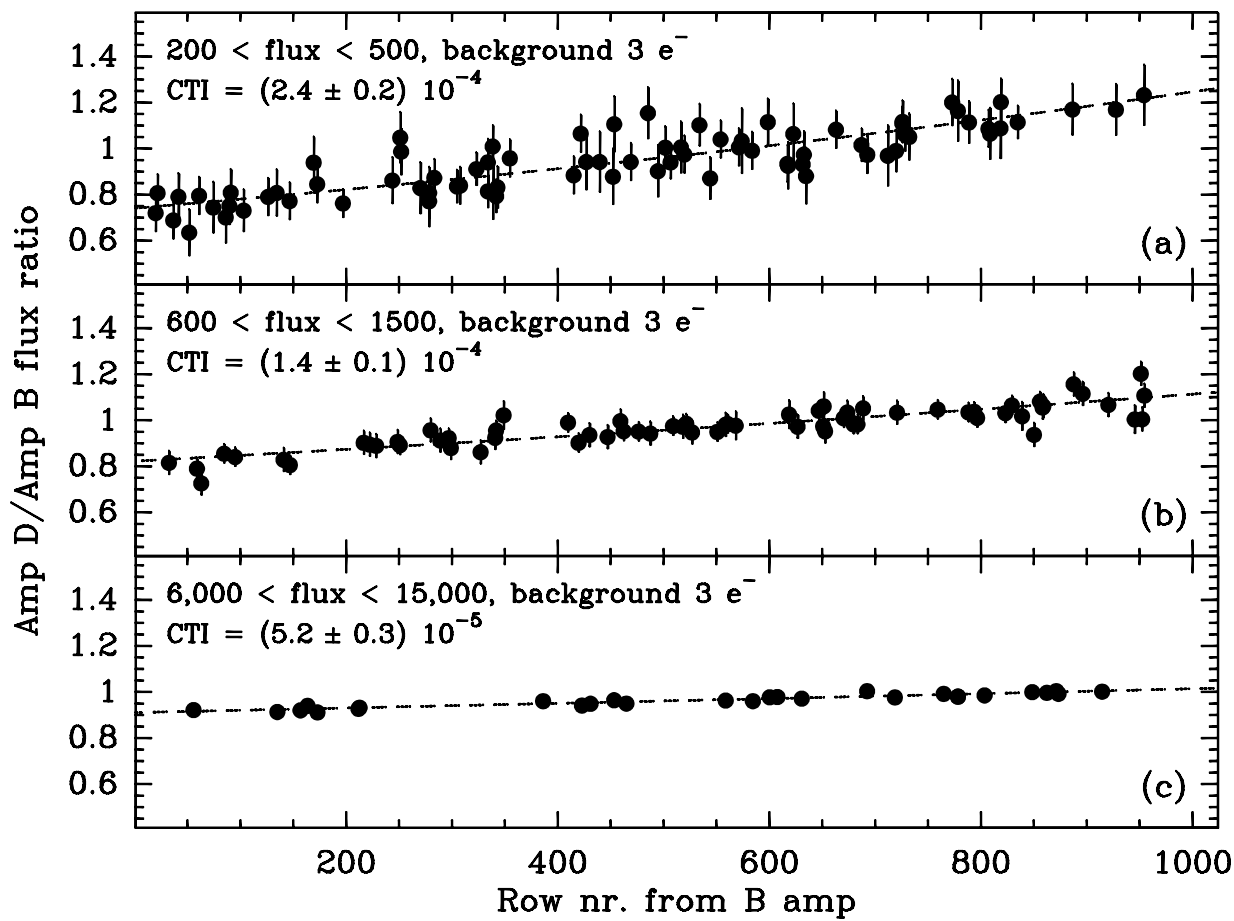}
\caption{Flux ratio results for the external imaging sparse field CTE test in 
  the parallel clocking direction, taken in October 2001 (after 4.7 years
  in orbit). The sky background was 3 e$^-$/pixel. The three panels
  correspond to measurements at different signal levels, as shown in the
  legends. In each panel, the ratio of the signals measured by 
  the two different amplifiers is plotted as a function of the Y position
  of the image along the CCD columns. The best-fitting simple CTE
  model (constant fractional charge loss per pixel transfer) as well as the
  corresponding CTI value are shown in each panel. The stellar fluxes shown
  here were measured using a circular aperture with a radius of 3
  pixels.\label{f:ESF_imagexample}}  
\end{figure}

Representative results on the parallel CTI for a short-exposure imaging
dataset acquired in October 2001 are shown in Figure~\ref{f:ESF_imagexample} in which
the observed flux ratio (amp D/amp B) vs.\ CCD row number is plotted for four
different ranges of stellar flux level per exposure. The expected CTE
behavior is clearly seen, with the closer readout amplifier systematically
measuring a higher stellar flux than the more distant amplifier. 
The drawn lines are fits of the CTE model we have been considering (i.e., Eq.\
\ref{eq:ctefunc}) to the data.  
Note that the charge loss incurred for parallel clocking through the image
area of the CCD can be quite substantial. For instance, panel (a) in
Figure~\ref{f:ESF_imagexample} shows that at that time, point
sources with a detected signal of a few hundred electrons suffer from
$\sim$\,25\% charge loss when located 1000 rows away from the read-out
amplifier. {\it Serial\/} CTI values were also determined (by reading out
with amplifier `C'), and found to be negligible for all practical purposes
(i.e., orders of magnitude smaller than the parallel CTI values, and
consistent with zero within the uncertainties).   
The default gain\,=\,1 setting (i.e., 1.0 e$^-$/DN) is used throughout.    

\begin{figure}
\plotone{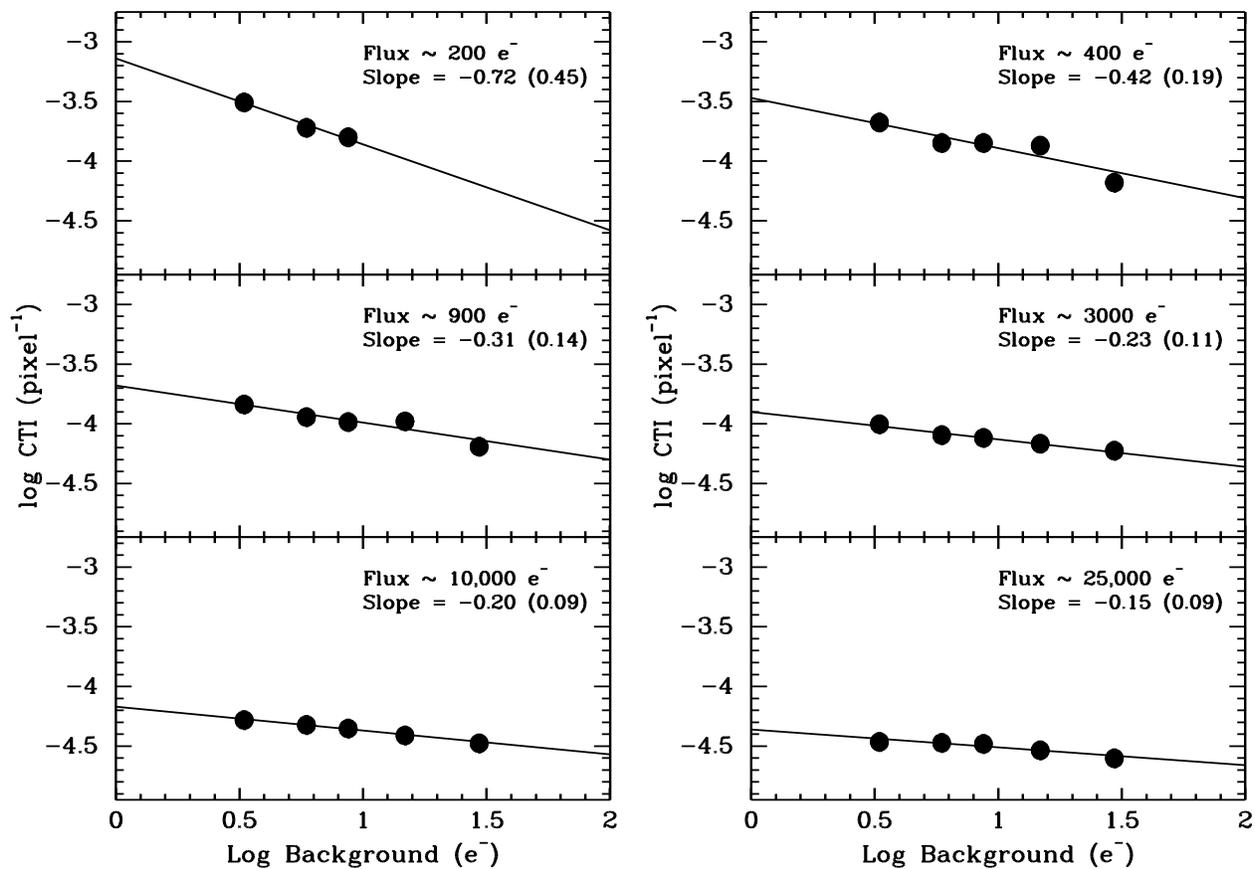}
\caption{Log-log plot of CTI vs.\ sky background level for six different
  signal levels, derived from the external imaging sparse field test taken
  after %2.8
  4.7 
  years in orbit. The flux levels, the slopes of the best-fitting line, and the
  latter's uncertainties (in parentheses) are mentioned in each panel. Note the
  systematic decrease of the CTI dependence on the sky background with
  increasing signal level.\label{f:ESF_skydep}}    
\end{figure}

Best-fitting parallel CTI values for different sky background levels are
plotted in Figure~\ref{f:ESF_skydep}. This plot illustrates two obvious trends
regarding the functional dependence of the CTE loss on source signal and 
background levels: {\it (i)\/} The CTI decreases with increasing
source signal level, and {\it (ii)\/} At a given source signal level,
the CTI decreases with increasing sky background levels. These
dependencies are also seen in investigations regarding CTE loss of
CCDs in the laboratory \citep*[e.g.,][]{hard+98}. 

A comparison of the fits to the data in the different panels of
Figure~\ref{f:ESF_skydep} shows that the slope of log\,(CTI) vs.\
log\,(background)  decreases systematically with increasing signal level. This
suggests {\it (i)\/} that sky background fills traps in the 
{\it bottoms\/} of the potential wells of the CCD, thus mostly ameliorating
the transfer of {\it small\/} charge packets, and {\it (ii)\/} a functional
dependence of the form 
\begin{equation}
 \mbox{CTI}\,(B,S) \propto \exp\left(-a\,(B/S)^b\right)\mbox{,} 
\label{eq:cti_B_S}
\end{equation}
where $B$ is the background level, $S$ is the signal level, and $a$ and $b$
are free parameters.  The substantial benefit of (only 
modest) sky background is good news for most science imaging observations
with the STIS CCD, which typically have longer exposure times than those used
for these tests and hence will not suffer from the large CTE effects
experienced during the low-background tests reported here. The functional form
of the dependence of CTI on signal and background levels will be
discussed further in Section~\ref{s:results}.

\section{Results} \label{s:results}

This section is organized as follows. We first analyse the internal sparse field
measurements to determine the time dependence of CTE loss and to
illustrate CTE effects at low signal levels when used with the
(supported but not commonly used) gain = 4 e$^-$/DN setting. We then determine
the functional forms to correct STIS CCD data for CTE  
loss as a function of observing mode (spectroscopy vs.\ imaging), 
signal level, 
and background level, and conclude with an examination of the
dependency of the CTE loss on aperture size in imaging mode. 

\subsection{Internal Sparse Field Data}

\subsubsection{The Time Dependence of CTE Loss and Centroid Shift} \label{s:timedep}

The internal sparse field test was the only CTE test conducted with the
STIS CCD in a uniform manner during both ground testing \citep[at a signal
level of $\sim$\,3400 e$^-$,][]{malu96} and in-flight 
operation (on an annual basis). Due to the combination of this long baseline
in time and the fact that CTI values measured from these data are of great
precision (as hundreds of columns are averaged together during
the analysis), we use these data to determine the time dependence of
the CTE loss. On the other hand, these data are not used to derive the
dependences of (spectroscopic) CTE loss on signal and background levels,
given the fact that these data sample only one (low) background level
per signal level, and this background level is different for the
different signal levels. 

\begin{figure}
\centerline{\includegraphics*[width=0.65\linewidth]{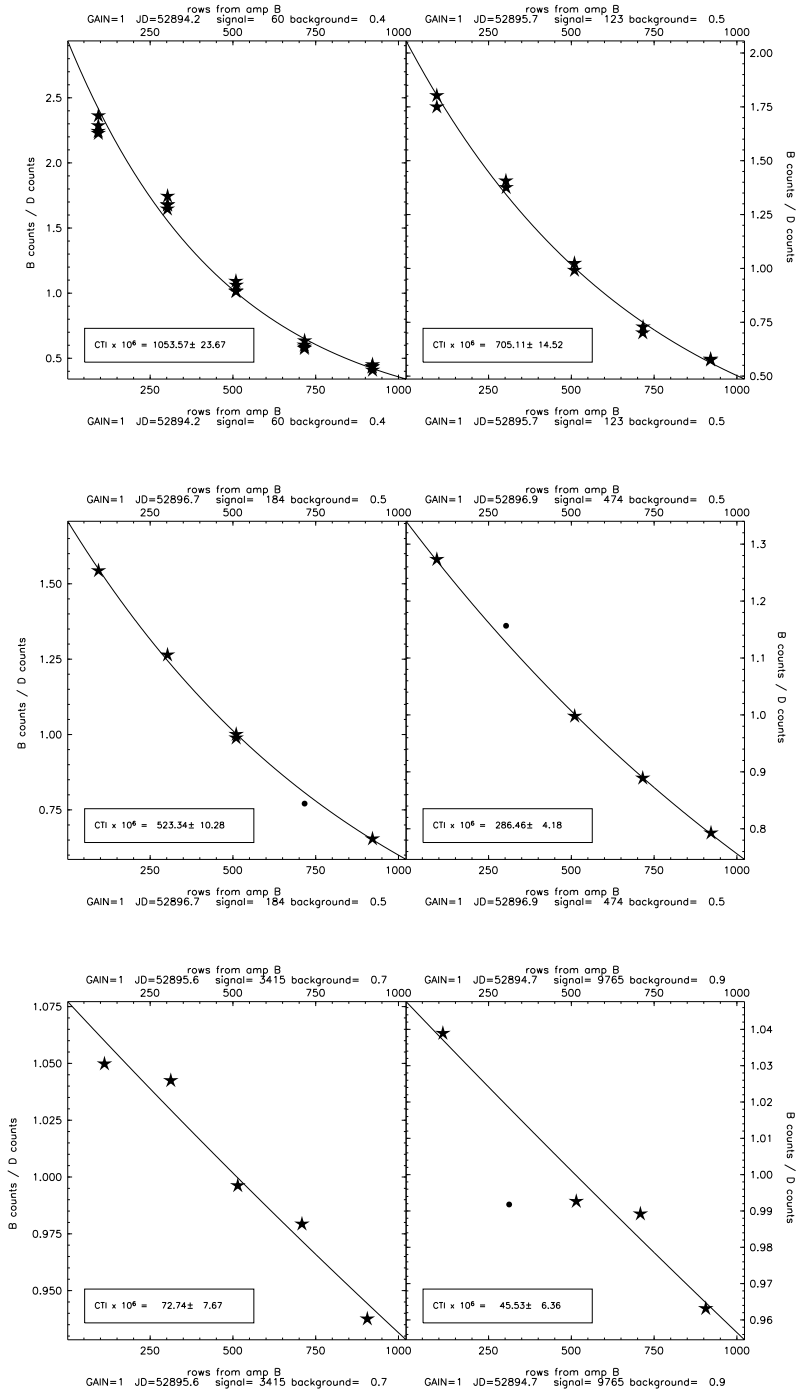}}
%\plotone{cte9620_gain1b.ps}
\caption{Flux ratio results for the 2002 epoch of the internal sparse field
  CTE test in  the parallel clocking direction. The six panels correspond to
  measurements at different signal levels, as indicated above and below each
  panel. In each panel, the ratio of the signals measured by the two 
  different amplifiers is plotted as a function of the mean Y position
  of the image along the CCD columns. The best-fitting simple CTE
  model (constant fractional charge loss per pixel transfer) as well as the
  corresponding CTI value are shown in each panel. Star symbols indicate
  measurements used in the fit and circles indicate rejected points. 
  \label{f:ISFexample}} 
\end{figure}

A $\chi^2$-minimization algorithm is used to compute CTI values for each
observing epoch and signal level. As outlined in Sect.~\ref{s:ISFtest}, the
observed ratio of the fluxes measured by the two amplifiers was fit to a
simple CTE model of constant fractional charge loss per pixel transfer,
allowing for $\kappa-\sigma$ clipping of outliers (the latter arise
occasionally during the short (0.3 s) exposures due to lamp intensity
fluctuations).  
Flux ratio results for the parallel internal sparse field test taken
after 6.5 years in orbit are presented in Figure~\ref{f:ISFexample}. As can be
seen there, the simple CTE model fits the data well. 

To derive the time dependence of the CTE loss, all measurements are first 
put on a uniform scale by normalizing them to one common background level $B'
= 0$. To do so, two corrections are required: First, 
the effect of the spurious charge in STIS CCD bias frames (Goudfrooij \&
Walsh 1997; this is further discussed below) is accounted
for by considering the total background $(B')$ to be 
the measured one $(B)$ plus the spurious charge. Second, the background
dependence of the CTI according to Eq.~\ref{eq:fin_specCTI} (which is
discussed in detail in Sect.~\ref{s:funcforms}) is taken into account. 
The time dependence is then derived by assuming a linear increase of CTI with
time, as was found earlier for other CCDs aboard {\it HST\/}
\citep[e.g.,][]{whit+99,dolp00}, i.e., by fitting the zero-background CTI values
to a function of the form:

\begin{equation}
{\rm CTI}\,(t) = {\rm CTI}_0 \; [1 + \alpha(t-t_0)],
\label{eq:timedep}
\end{equation}

\noindent with $t$ in years and $t_0 = 2000.6$, the approximate midpoint in
time of in-flight STIS observations. The conversion between ($t-t_0$) and the
modified Julian date MJD (which is provided by keywords {\tt TEXPSTRT} and
{\tt TEXPEND} in the science header of each STIS observation) is given by 
$t - t_0 = ({\rm MJD}-51765)/365.25$.  

\begin{figure}[tb]
\centerline{%\psfig{file=f6.eps,width=16.cm}
\includegraphics[width=\textwidth]{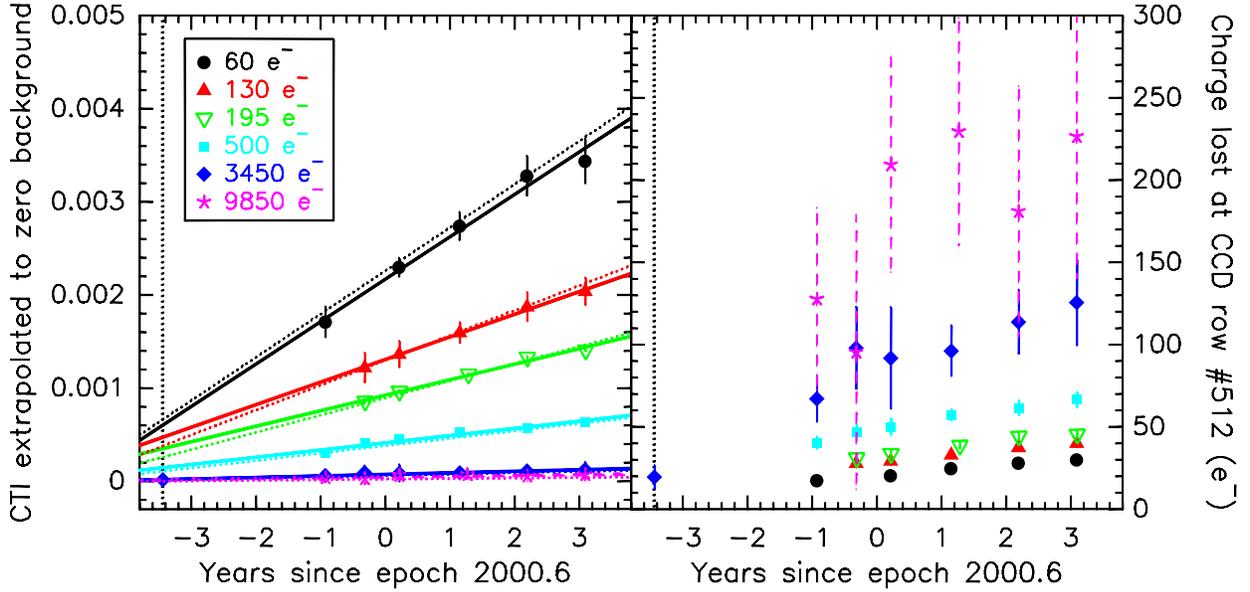}
}
\caption{{\it Left panel:\/} CTI normalized to zero background for gain\,=\,1 as
  a function of time and signal level, derived from the internal sparse field
  test. Symbols associated with the data of individual signal levels
  (corrected for CTI) are indicated in the legend. The solid lines represent
  linear fits to the data (one for each signal level), while the dotted lines
  represent the result of applying Eq.~\ref{eq:fin_specCTI} to the data (one
  line per signal level group). 
  {\it Right panel:\/} Absolute charge lost due to CTI for an object at the
  central row of the STIS CCD as a function of time and signal level. Symbol
  types are the same as in the left panel. The epoch of {\it HST\/} Servicing
  Mission 2 (during which STIS was installed on {\it HST}) is depicted as a black
  dotted vertical line.\label{f:ISF_CTIevol_g1}}
\end{figure}

CTI values derived as mentioned above for the parallel internal sparse field
test taken at different epochs are plotted in Figure~\ref{f:ISF_CTIevol_g1}. 
In-flight CTE degradation from a pre-flight starting point of low CTI is
apparent. Typical CTE behavior is observed as a function of
signal level: The {\it fractional\/} charge loss (which is proportional to
CTI) drops with increasing signal level, while the {\it absolute\/} level of
charge loss increases.   

Results for the time-dependence fit for gain\,=\,1 are shown in
Figure~\ref{f:ISF_CTIevol_g1} and Table~\ref{t:alphaval}. The functional fit to the
data (shown as solid lines) is good, and the derived values for $\alpha$ in Eq.\ 2 are
consistent with one another (within the uncertainties) for all signal levels 
measured. 
After weighting the $\alpha$ values of the individual fits by their inverse
variance (i.e., RMS$^{-2}$), the weighted average time constant was found to be
$\alpha = 0.205 \pm 0.006$, where the uncertainty represents the mean
error of the mean. 

\begin{figure}
\centerline{%\psfig{figure=f7.eps,width=0.8\linewidth}
\includegraphics[width=\textwidth]{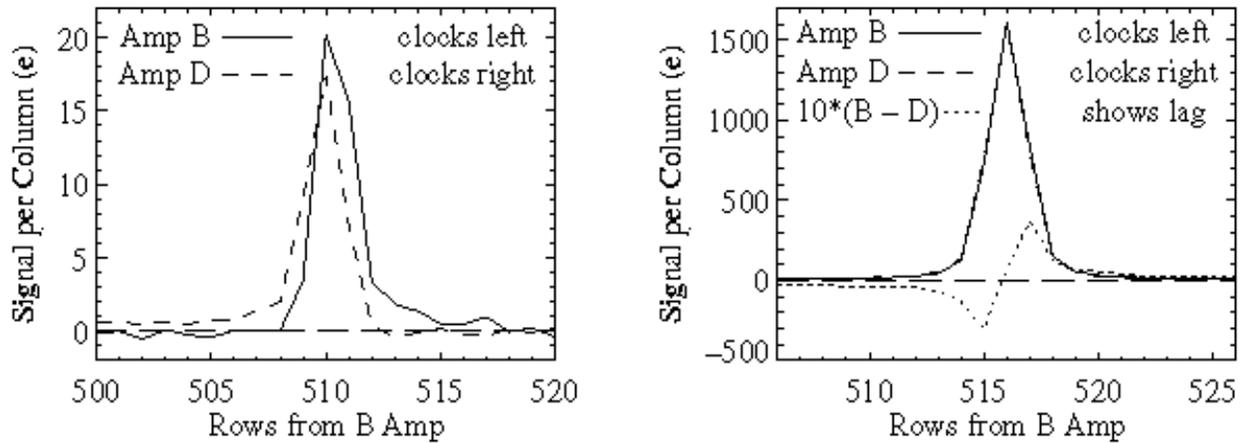}
}
\caption{Comparison of line profiles obtained
  when clocking a given exposure in opposite directions in the
  internal sparse field test. At low signal levels (left panel),
  charge trailing and centroid shift are obvious. Even at high signal levels
  (right panel), differences in the line profile are systematic
  and measureable (the dotted line depicts the difference
  profile multiplied by 10). Figure reproduced from Kimble,
  Goudfrooij, \& Gilliland (2000).}
\label{f:ISF_lineprof1}
\end{figure}

\begin{figure}
\centerline{\includegraphics*[width=0.9\linewidth]{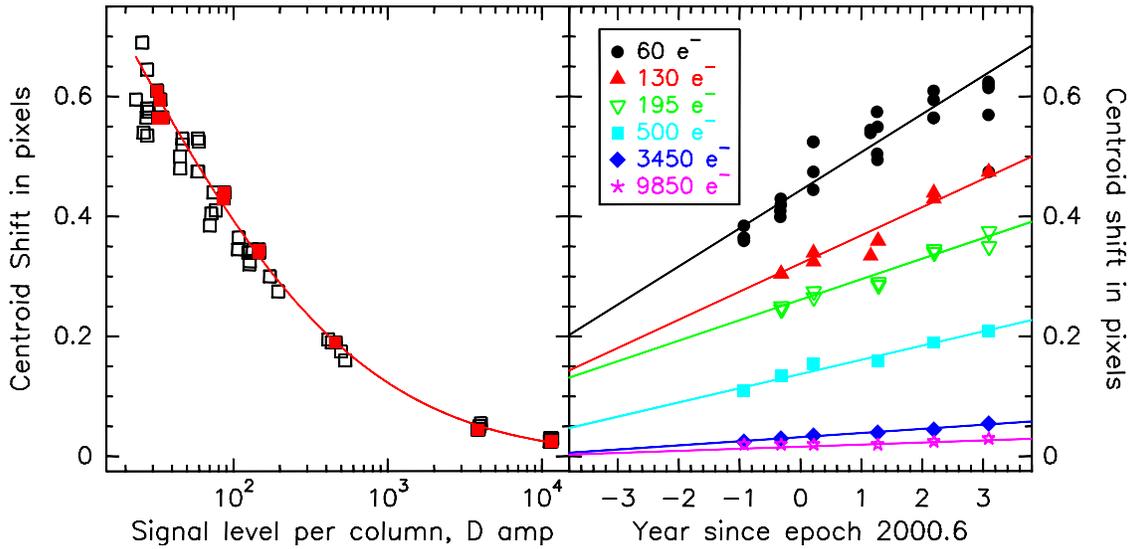}}
\caption{{\it Left panel}: The centroid shift (in
  unbinned CCD pixels) as a function of signal level as read out by
  the D amplifier for the gain\,=\,1
  observing block of the internal sparse field test in October 2002, $\sim$\,5.5
  years after STIS installation. Centroid shifts for the central location on the CCD
  are shown in filled squares, and a least-squares fit to the latter
  is shown by the solid line. {\it Right panel}: The centroid
  shift for the central location on the CCD as functions of time and
  signal level as read out by the D amplifier. Both the data and the
  corresponding linear fits are plotted. Symbols associated with
  individual signal levels (corrected for CTI) are indicated in the legend.  }  
\label{f:ISF_lineprof2}
\end{figure}

The effects of charge trapping and release {\it within\/} the 7-row
aperture are seen by examining the line profiles and
centroids. Comparisons of the average line profiles seen for the
opposing readout directions are shown in 
Figure~\ref{f:ISF_lineprof1} for two of the signal levels used. At 60 electrons
per column, the charge trailing and centroid shift are obvious. Even at the
higher signal level, with much lower CTI, the magnified difference between the 
two observed profiles shows that the centroid shift is systematic and
measurable. The measured centroid shifts (defined as half the
difference between the profile centroids as measured by the B vs.\ the D 
amplifiers) after 5.5 years in orbit are
plotted in the left panel of Figure~\ref{f:ISF_lineprof2}, along with
a least-square fit to the centroid shifts measured at the
central position on the CCD as a function of measured gross signal
level $G$ as read out by the default amplifier D. 
The right panel of Figure~\ref{f:ISF_lineprof2} depicts the {\it
  evolution\/} of the centroid shift (at the central location on the
CCD) due to CTE effects as a function of time. A comparison with
Figure~\ref{f:ISF_CTIevol_g1} shows that the
centroid shifts increase in time with  rates that are quite similar
to those measured for the CTI values themselves. 

\begin{figure}
\centerline{\includegraphics*[width=0.6\textwidth]{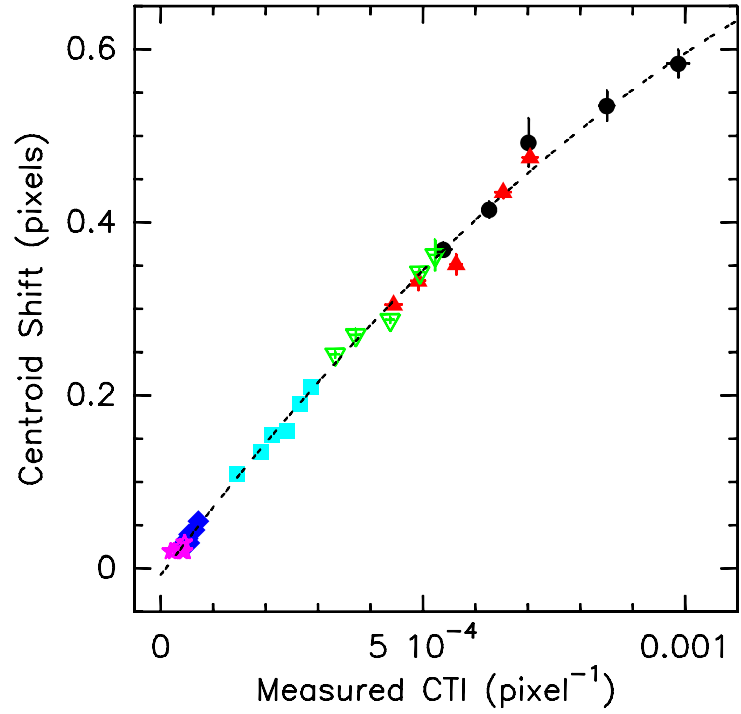}}
\caption{The centroid shift at the central row of the CCD as a function of
  CTI for all gain\,=\,1 datasets of the internal sparse field test. Symbols as in
  Figure~\ref{f:ISF_lineprof2}. Note the very
  tight relation between the two properties. A least-squares fit to the data
  (cf.\ Eq.~\ref{eq:shift_from_cti}) is shown by the dashed line.}  
\label{f:cti_vs_shift}
\end{figure}

All the above findings indicate that CTE effects cause a fractional
loss of signal of which the amplitude is directly related to the size
of the associated centroid shift, anywhere on the CCD. This is
illustrated further in Figure~\ref{f:cti_vs_shift}, which reveals an extremely 
tight relation between measured CTI and centroid shift. The solid line
depicts the least-square fit to this relation,
\begin{equation}
\mbox{Centroid Shift [pixels]} = 0.081 \, \left(
  \frac{\mbox{CTI}}{10^{-4}} \right) \,
 - 0.002 \, \left( \frac{\mbox{CTI}}{10^{-4}} \right)^2 
\label{eq:shift_from_cti}
\end{equation}
which has an RMS error of only 0.01 pixels. (Note that positive centroid shifts in this
equation indicate shifts toward {\it smaller\/} Y coordinate values in the
nominal STIS coordinate system.) This relation should prove
useful to spectroscopic programs for which both accurate flux calibration 
{\it and\/} astrometry is important.

\begin{figure}
\centerline{\includegraphics*[width=0.95\textwidth]{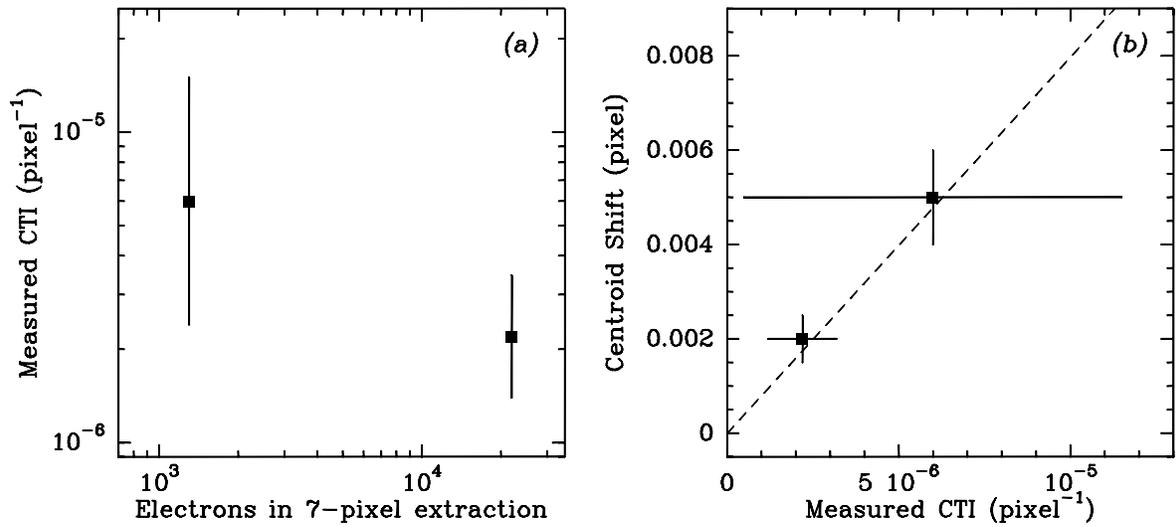}}
\caption{Panel (a): Serial CTI as a function of signal level for the 1999
  epoch of the internal sparse field test. Panel (b): The centroid shift at the central
  column of the CCD as a function of CTI for the same data. The relation
  between CTI and centroid shift found for the parallel CTI
  (Eq.~\ref{eq:shift_from_cti}), as shown in Figure~\ref{f:cti_vs_shift}) is
  shown by the dashed line.}
\label{f:isf_serialcti}
\end{figure}

Similar measurements have also been carried out in the serial clocking 
direction by projecting vertical slit images at several positions along the CCD
rows and alternating between amplifiers `C' and `D' (cf.\ Figure~\ref{f:ccdarch})
for readout. The resulting charge loss is {\it much\/}
smaller than for the parallel case and stays below 1\% for 1024 pixel
transfers, even for the lowest signal level measured. This significant
difference between the CTE loss 
experienced in the two clocking directions is mainly due to the difference in
clocking time, which is a factor $\sim$\,30 faster in the serial
direction for the STIS CCD \citep[22 $\mu$s pixel$^{-1}$ in serial vs.\ 640
$\mu$s pixel$^{-1}$ in the parallel direction; see also][]{brisalex02}. 
Centroid shifts in the serial clocking direction are also very small but 
still well determined, due to the statistical precision offered by
averaging over hundreds of rows. Interestingly, the serial CTI values and centroid
shifts are consistent with the relation found above in
Eq.~\ref{eq:shift_from_cti}. This is illustrated in
Figure~\ref{f:isf_serialcti}. 
Given the negligible CTE loss experienced by the STIS CCD in the
serial clocking direction, the remainder of this paper only addresses
the {\it parallel\/} CTE loss.  
 
\subsubsection{CTE Effects at Low Signal Level: Issues for the Gain\,=\,4
  Setting}  \label{s:lowsig}  

\begin{figure}
\plotone{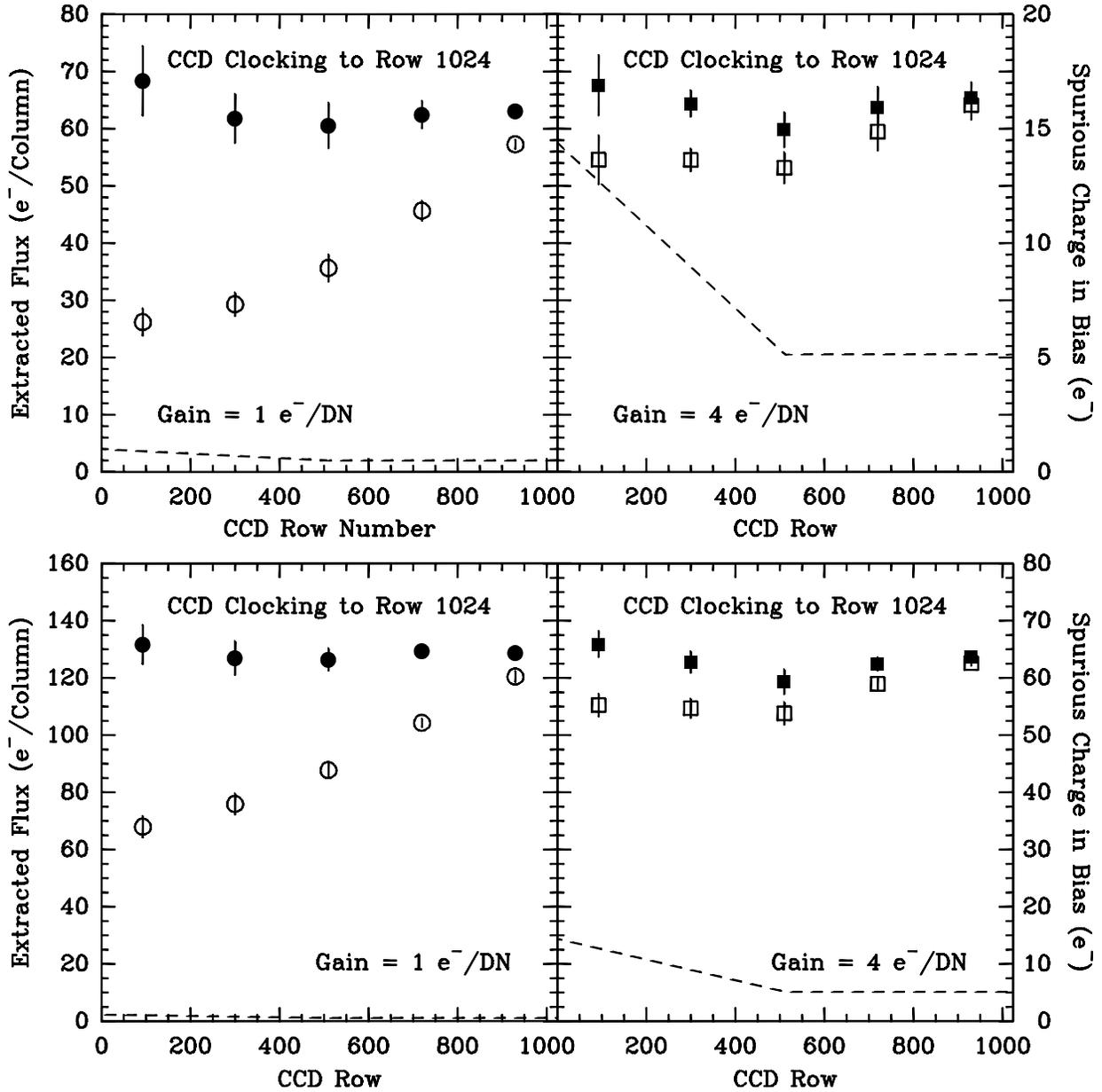}
\caption{Results of the two lowest-signal cases of the 2002 epoch data of the
  internal sparse field test in the parallel clocking direction, for the two gain
  settings (1 and 4 e$^-$/DN). The average signals measured at 
  the five different positions along the CCD columns are shown as symbols
  (open circles for gain\,=\,1 (left panels), open squares for gain\,=\,4
  (right panels)). Extracted signal levels are shown on the left-hand
  ordinate. The dashed lines depict the `background' level due to spurious 
  charge in STIS CCD superbias frames, averaged over all columns. Superbias
  intensity levels (in e$^-$ per pixel) are shown on the right-hand
  ordinates. Note the strong effect of local `background' to the CTI for these low
  signal levels, especially for gain\,=\,4 data. The latter show a much
  lower CTI than the gain\,=\,1 data, due to a significantly higher amount of
  spurious charge in gain\,=\,4. The filled symbols depict the signal levels after
  correction for CTI using Eqs.~\ref{eq:fin_specCTI} and \ref{eq:implement}.
  \label{f:ISFlowsignal}}   
\end{figure}

As mentioned above in Section~\ref{s:ESFtest} and shown in
Figure~\ref{f:ESF_skydep}, there is a strong dependence of the CTE loss on
background level, especially for low signal levels. In fact, this effect is
significant enough for the structure of STIS CCD bias frames to cause an
apparent non-linearity in the CTI values derived from the internal 
sparse field tests at the lowest signal levels. This is illustrated 
in Figure~\ref{f:ISFlowsignal}. The data are shown in two panels per
signal level, one for each supported CCD gain setting (1.00 and 4.08 e$^-$/DN,
hereafter referred to as gain\,=\,1 and gain\,=\,4 respectively). The
open symbols depict actual flux measurements (flux levels are indicated on the
left ordinate) of the epoch 2002 data of the internal sparse field test. The flux
measurements with the B and D amps were averaged together for this
purpose, after having flipped the CCD row numbers for the B-amp measurements
around the central row ({\#}512) in order to coincide with the D-amp rows. The
error bars indicate the measurement repeatability. 
The gain\,=\,4 setting features a CTE-induced loss that is
significantly smaller than that for the gain\,=\,1 setting. The dashed
lines depict the structure of STIS bias frames (after subtraction of the overscan
vector) in the appropriate gain setting along CCD columns. The STIS CCD suffers
from accumulative {\it spurious charge\/}, which causes a row-dependent amount
of extra ``background'' that is added during readout
\citep[see][]{gouwal97}. The amount of added spurious charge for 
the gain\,=\,4 setting is much larger than that for gain\,=\,1 (due to a
difference in parallel clocking voltages employed), and the
slope in the spurious charge level across the lower half of the CCD is much
steeper for gain\,=\,4 than for gain\,=\,1 bias frames. This is the main cause
of the difference in low-signal level CTE behavior between the two gain
settings. Jumping ahead, the filled symbols 
in Figure~\ref{f:ISFlowsignal} depict the result of the application of
the CTE correction formula (Eq.~\ref{eq:fin_specCTI}) derived in
Section~\ref{s:funcforms} below. The corrected fluxes are uniform across the
CCD to within $\sim$\,1 $\sigma$ for gain\,=\,1, while gain\,=\,4 data at low
signal levels are still somewhat affected by the slope of the spurious charge in
the bias frames. Hence, the accuracy of the CTE corrections discussed below
formally applies only to gain\,=\,1 data (and gain\,=\,4 data at signal levels
$\ga 2000$ e$^-$)\footnote{The main purpose of the gain\,=\,4 setting of the
STIS CCD is to enable very high signal-to-noise observations, using signal
levels up to the intrinsic full well of the CCD, at the cost of 
a higher read noise than for gain\,=\,1 \citep[see Chapter 7
in][]{kimq+04}. This behavior at low signal levels is therefore not
detrimental for {\it typical\/} data taken in gain\,=\,4.}. 

\subsection{Functional Forms for CTE Corrections}   \label{s:funcforms}

Since spectroscopy and imaging involve inherently different detector illumination
patterns, CTE measurements are expected to show different dependencies on
signal level and background level for the two observing modes. For instance,
a given total signal level within a certain aperture size in imaging mode
contains a substantial range of signal levels {\it per CCD column}, and 
the charge in each of those CCD columns suffers from a different percentage of
CTE loss. With this in mind, we chose to determine functional forms to
characterize the CTE loss of the STIS CCD separately for spectroscopy
and imaging modes.   

\subsubsection{Spectroscopy Mode} \label{s:spec_cti}
 
\paragraph{Dependence on Signal and Background Level} 

\begin{figure}
\plotone{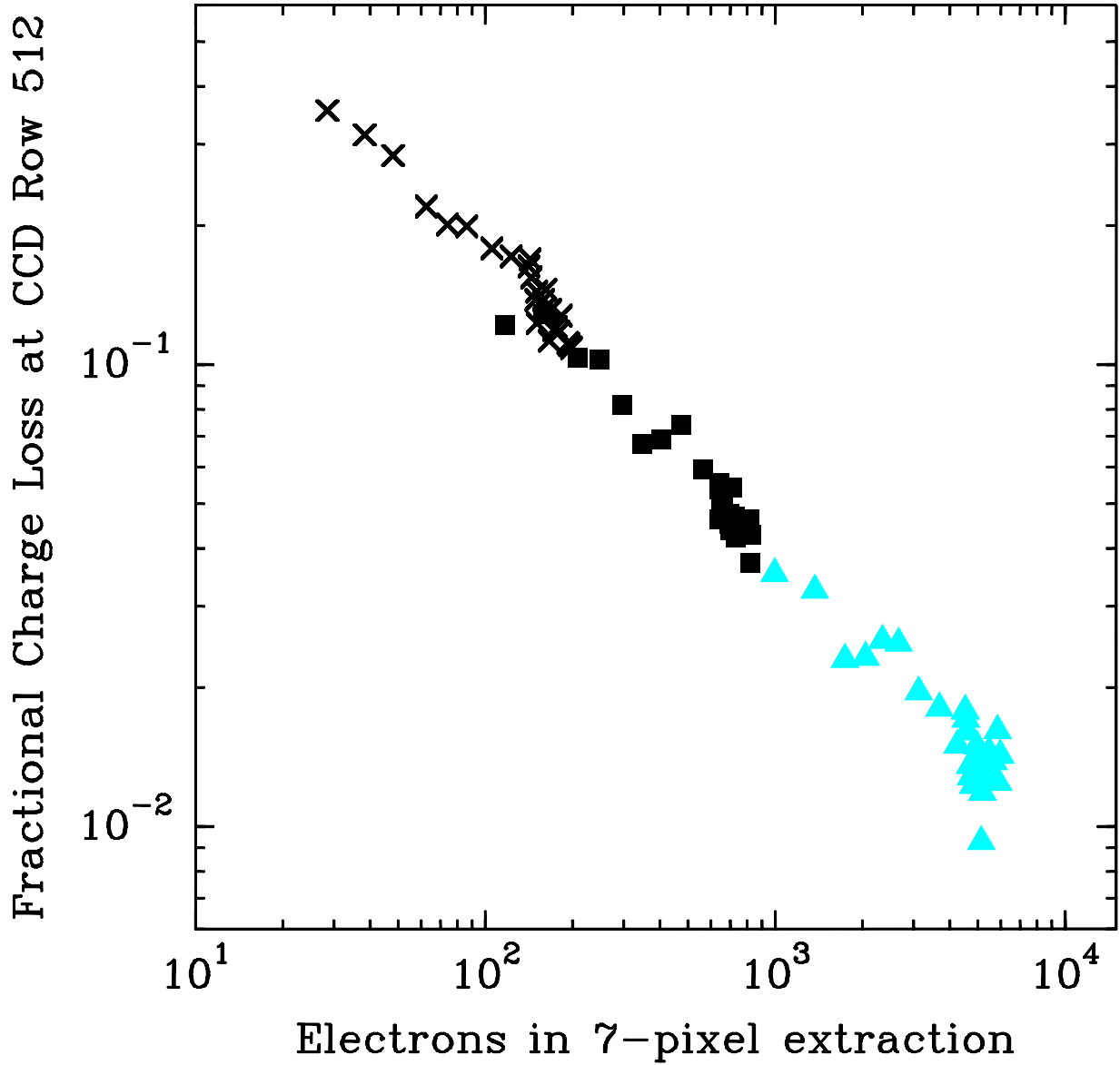}
\caption{Fractional charge loss at the central row of the CCD (i.e.,
  (1\,$-$\,CTI)$^{512}$) vs.\
  extracted gross signal level for three G230LB wide-slit 
  spectra of flux standard stars observed at the center of the STIS
  CCD. Crosses, squares, and triangles represent values for spectra {\tt
    o8v204030}, {\tt 08v2040e0}, and {\tt o6ig10010}, respectively (see
  Table~\ref{t:spectra}). The CTI values were normalized to correspond to
  epoch 2000.6 
  (see Sect.~\ref{s:timedep}). Note the linear dependence of
  log\,(CTI) on log\,(signal level).\label{f:g230lb_cti}}     
\end{figure}

\begin{figure}
\includegraphics[width=\textwidth]{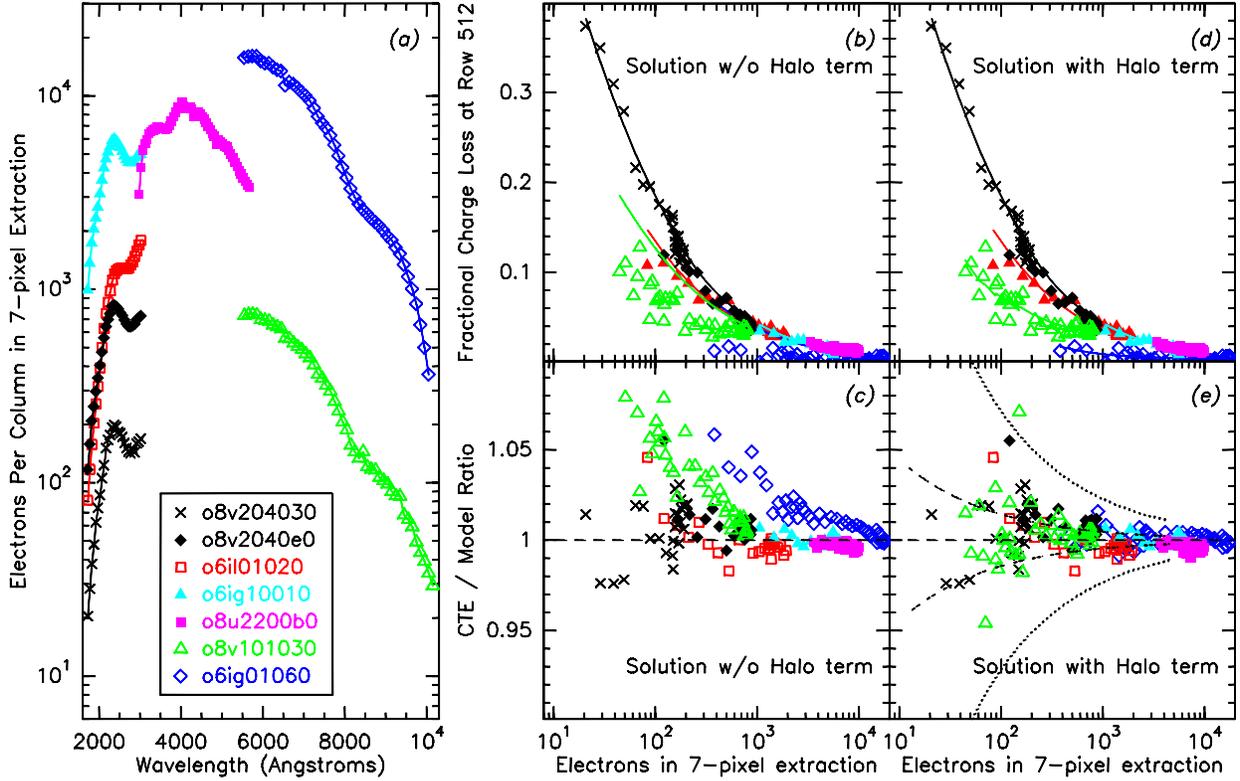}
\caption{{\it Panel (a):\/} Smoothed flux standard star spectra used to
  determine the functional form of the CTI of the STIS CCD in spectroscopic
  mode. The legend links the symbols with the STIS rootnames of the
  individual spectra, as listed in Table~\ref{t:spectra}.  
  {\it Panel (b):\/} Fractional charge loss at the central row of the CCD
  vs.\ gross signal level within the default 7-pixel extraction
  box. Symbols represent measured CTI values for the spectra shown in panel
  (a), normalized to epoch 2000.6 
  (see Sect.~\ref{s:timedep}). The smooth curves
  represent the predictions of the best-fit CTI model {\it without\/} `red
  halo' term [i.e., Eq.~\ref{eq:cti_B_G}] for those data. Symbol types
  are the same as in panel (a). 
  {\it Panel (c):\/} The ratio of measured CTE values and the model predictions
  shown in panel (b) vs.\ gross signal level.  
  {\it Panel (d):\/} Same as panel (b), but now using the final 
  CTI model including a `red halo' term [i.e., Eq.~\ref{eq:fin_specCTI}]. 
  {\it Panel (e):\/} Same as panel (c), but now using the final 
  CTI model including a `red halo' term. For reference, the dashed lines
  represent the uncertainty due to Poisson noise associated with the binned
  spectra shown in panel (a) as a function of signal level, while the dotted
  lines represent the Poisson error associated with a resolution element (2
  pixels along the dispersion) of unbinned spectra. 
  }  
\label{f:cti_specplot}
\end{figure}

To constrain the dependence of spectroscopic CTE loss on signal and background
levels, we utilize a variety of spectra of spectrophotometric standard stars
taken during the last few years of STIS operations, when CTE effects were most
significant. The standard star spectra used to characterize the CTE 
effects in this report are listed in Table~\ref{t:spectra} along with their
measured signal and background levels. All spectra used a 2$''$ wide slit
to render slit losses negligible. Spectra are calibrated using {\sc
  calstis} {\it without\/} applying any CTE correction, and extracted using
traces measured from the spectra themselves (see \citet*{dres+06} regarding the
evolution of spectral traces with STIS). CTI values for spectra of DA white
dwarf flux standards GD\,71 and  LDS\,749B taken using the G230LB grating are
derived from the ratio of their measured fluxes by those measured from G230L
spectra. The (time-dependent) sensitivity calibration for the G230L mode
(which uses the NUV-MAMA detector of STIS (a microchannel-plate based
photon-counter, not a CCD) and hence does {\it not\/} suffer from CTE loss)
is very well established and accurate to subpercent level \citep{stys+04}, 
while the sensitivity calibration for the CCD spectral modes was
established from flux standard star spectra with signal levels high enough
($\ga 2 \times 10^4$ e$^-$/extraction) to render CTE effects
negligible. 
Figure~\ref{f:g230lb_cti} shows the CTI values obtained this way from the
G230LB spectra of flux standard stars GD\,71 and G191$-$B2B listed in
Table~\ref{t:spectra}, which share a common background level. Prior to
plotting, these CTI values were first normalized to epoch 2000.6 (see
Sect.~\ref{s:timedep} above). 
The charge loss for low-signal spectra is already quite
significant in mid-2000, rising above 10\% below an extracted signal level
of $\sim$\,200 e$^-$ for spectra taken at the nominal central position on the
CCD.  
Figure~\ref{f:g230lb_cti} shows that the logarithm of CTI scales roughly
linearly with the logarithm of signal level, i.e.: 
\begin{equation}
 \mbox{CTI}\,(G) \propto \alpha\,G^{-\beta}\mbox{,}
\label{eq:cti_G}
\end{equation}
where $G$ is the (gross) signal level and $\alpha$ and $\beta$ are free
parameters. For the G430L and G750L spectra in Table~\ref{t:spectra}, CTI
values are  derived by comparing the observed fluxes with pure hydrogen white dwarf
models, as detailed in \citet*{bohl+97}. 
The range of extracted signal
levels covered by the spectra used here is $\sim$\,20 -- 16,000 e$^-$ pixel$^{-1}$,
which should encompass all spectroscopic observations for which CTE
loss is an issue. Smoothed versions of the spectra are shown in panel (a) of
Figure~\ref{f:cti_specplot}. 

For the expected basic functional form of the dependence of the CTE
loss on signal and background levels we combine Eqs.\
\ref{eq:cti_B_S} and \ref{eq:cti_G}:  
\begin{equation}
 \mbox{CTI}\,(B,G) = \alpha\,G^{-\beta} \;
 \exp\left(-\gamma\,(B'/G)^\delta\right)\mbox{,}
\label{eq:cti_B_G} 
\end{equation}
where $B'$ is the {\it total\/} background signal level (i.e., the sum of the sky
background $B$, the dark current, and the spurious charge in STIS bias
frames which is 0.5 e$^-$ pixel$^{-1}$ for gain\,=\,1 and 5.0 e$^-$
pixel$^{-1}$ for gain\,=\,4),  
$G$ is the gross signal level, and $\alpha$, $\beta$, $\gamma$ and $\delta$
are free parameters.  The values for $B$ and $G$ are readily obtained from the
output of the {\sc x1d} routine within {\sc calstis} to extract 1-D spectra
\citep{mcgr+99}.  The background spectrum in {\sc x1d} is smoothed
in the spatial and dispersion directions prior to subtraction; this smoothed
version is the $B$ used in Eq.\ \ref{eq:cti_B_G}.  

An attempt to fit the function in Eq.~\ref{eq:cti_B_G} to the CTI values
measured for the spectra in Table~\ref{t:spectra} showed that the CTI values
of the G230LB and G430L spectra can be fit very well this way. However, the
CTI values measured for the red end of the wavelength coverage of the G750L
spectra are systematically found to be below the predictions of this fit. This
is illustrated in panels (b) and (c) of Figure~\ref{f:cti_specplot} which show
the best fit of Eq.~\ref{eq:cti_B_G} to the G230LB and G430L data and the
residual percentage of flux error after correcting the data for the CTE loss 
associated with that fit, respectively. In the following we attempt to
correct for this problem by parametrizing the background parameter in
Eq.~\ref{eq:cti_B_G} in terms of the influence of the signal in the wings of
the point-spread function (PSF) to the filling of charge traps during the CCD
readout procedure.   

\paragraph{The Impact of the ``Red Halo'' of the PSF of the STIS CCD}

The PSF of the STIS CCD features broad wings at wavelengths $\ga 8000$ \AA\
\citep[e.g.,][]{leiboh97}, the width of which increases strongly with
increasing wavelength. This ``red halo'' is believed to be due to scatter
within the CCD mounting substrate which becomes more pronounced as the silicon
transparency increases at long wavelengths. The effects of the red halo are
significant, particularly redward of 9500\,\AA\ where the default 7-pixel
extraction box captures only $\la 70$\% of the light in the PSF
\citep{leiboh97}. 

\begin{figure}
\centerline{%\psfig{figure=f13.eps,width=0.9\textwidth}
\includegraphics[width=\textwidth]{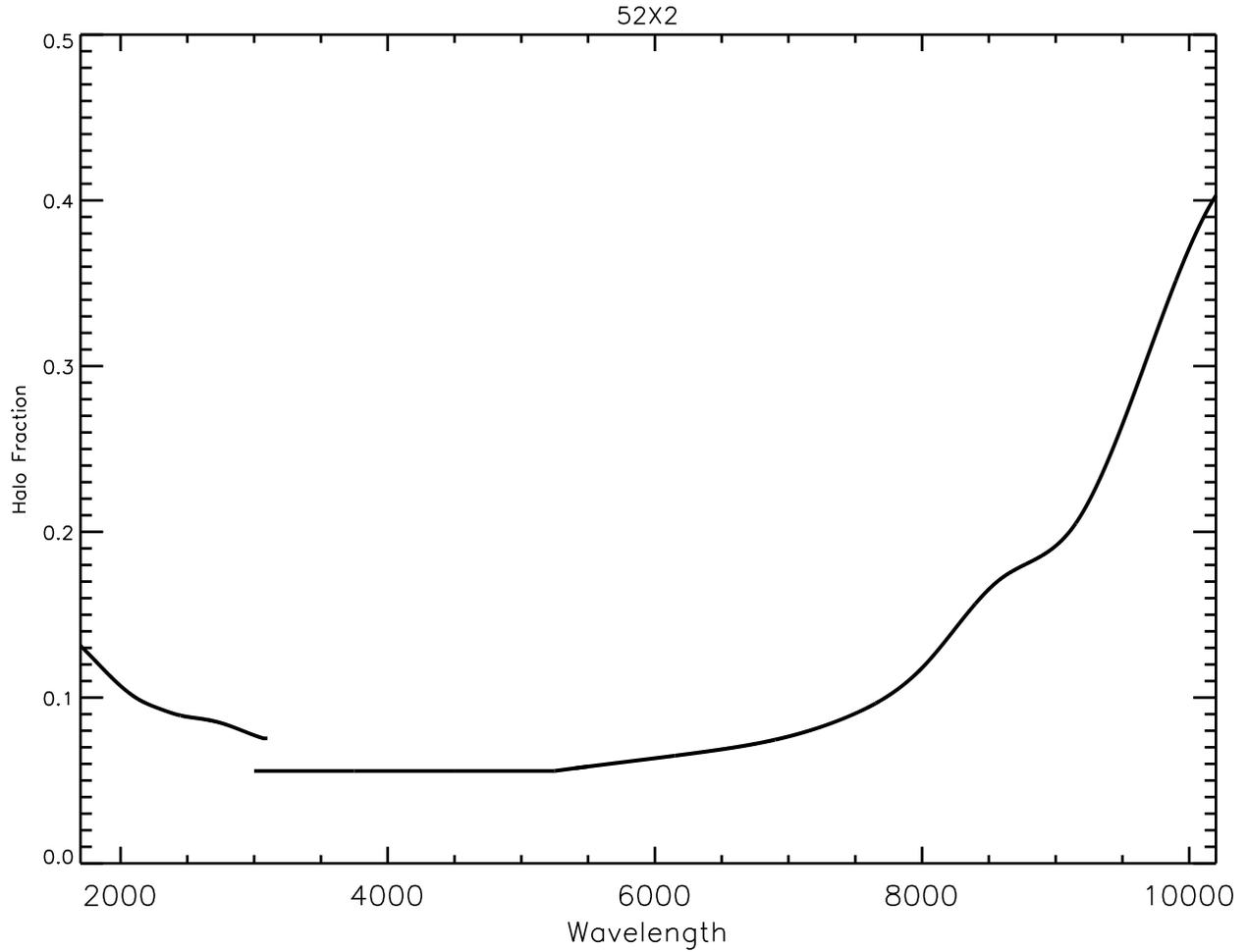}
}
\caption{Parameter $H$ in Eq.~\ref{eq:fin_specCTI}: The fraction of the light
  in the PSF above the default 7-pixel extraction box as a function of
  wavelength for the STIS CCD grating modes.  Note the discontinuity near
  3000\,\AA, at the boundary of the wavelength ranges covered by the G230LB
  and G430L gratings. This is likely due to the presence of a Lyot stop in the
  G430L and G750L modes which is absent in the G230LB mode \citep[see
  also][]{dres+06,prof06}.}  
\label{f:Hvalues}
\end{figure}

This extended halo is likely to have a significant effect on the CTE loss 
experienced by the signal within the default 7-pixel extraction box, since 
the charge induced by the halo signal that is clocked out {\it before\/} the
signal of the source spectrum (i.e., the halo signal on the side of the
read-out amplifier) should act effectively as `background' in filling
charge traps. Note that the red halo signal is not included in the
background term $B'$ in Eq.~\ref{eq:cti_B_G}, since the background
spectrum used within {\sc calstis/x1d} is measured far away from the spectrum
location\footnote{300 unbinned CCD pixels away by default.}. To
improve this situation, we separate the background term into two distinct 
terms, $B'$ (as before) and a new term $H$ which contains the fraction of PSF
signal between the default 7-pixel extraction box and the read-out
amplifier. Values for $H$ as a function of wavelength are derived from
existing {\sc calstis} reference files (namely from the Photometric
Correction Tables\footnote{{\tt *\_pct.fits}, listed in data header keyword
  PCTAB.}). We plot these $H$ values in Figure~\ref{f:Hvalues}. While $H$
is non-negligible at any wavelength, the spatial extent of the
PSF beyond the default extraction box is only a few CCD pixels below
$\sim$\,8000\,\AA. Hence, low values of $H$ do {\it not\/} necessarily lower
the CTE loss significantly. This issue is accounted for in two ways: {\it
  (i)\/} by subtracting a certain minimum threshold value from the measured
value of $H$ (i.e., parameter $\eta$ in Eq.~\ref{eq:fin_specCTI} below), and
{\it (ii)\/} by neglecting the halo term for gratings other than G750L or G750M. 

Taking all of the above into account, the full functional form of the CTI
algorithm considered for spectroscopic modes is 
\begin{eqnarray}
\mbox{CTI}_{\it SP} & \!=\! &  \alpha\, G^{-\beta} \; (\gamma(t - 2000.6) + 1) \;
\exp \left(-\delta\left[\frac{B'+\epsilon H'}{G} \right]^\zeta \, \right) 
\label{eq:fin_specCTI} \\
\mbox{where $H'$} & \!=\! & 
 \mbox{max}(0.0,(H-\eta))\times \mbox{{\it Net}.} \nonumber 
\end{eqnarray}
({\it Net\/} = $G-7B$, the net counts in the extracted spectrum.) 
Initial estimates of the values of parameters $\alpha$ through $\eta$
and their uncertainties were made using bootstrap tests (except for the CTI
time constant $\gamma$, which was fixed at 0.205 as discussed in
Sect.~\ref{s:timedep} above). Particular attention was given to
achieving a suitable fit for the full range of signal and background levels
sampled. A robust fit parameter was then minimized using a 
non-linear minimization routine from {\it Numerical Recipes\/} (Press et al.\
1992). 
Best-fit values of the parameters $\alpha$ through $\eta$ are listed in
Table~\ref{t:speccticoeff}.   

The quality of this parametrization of the spectroscopic CTE loss is
illustrated in Figures~\ref{f:ISF_CTIevol_g1} and
\ref{f:cti_specplot}. The dotted lines in the left panel of
Figure~\ref{f:ISF_CTIevol_g1} show the CTI values predicted by
Eq.~\ref{eq:fin_specCTI} for the internal sparse field data at
the different gross signal levels (and $B' = 0$). It can be seen that
Eq.~\ref{eq:fin_specCTI} fits those data within the uncertainties\footnote{The small
offset for the 60 e$^-$/extraction data is due to the effect of the small
slope of the spurious charge level in the lower half of STIS CCD bias frames,
as discussed in Sect.~\ref{s:lowsig} and illustrated in
Figure~\ref{f:ISFlowsignal}.}.  Figure~\ref{f:cti_specplot} shows the fit of
Eq.~\ref{eq:fin_specCTI} to the standard star spectra, both with and without
the $H$ term mentioned above. A comparison of panels (c) and (e) of
Figure~\ref{f:cti_specplot} in particular clarifies that the inclusion of the
$H$ term yields a significantly better correction for the red end of G750L spectra. 
Overall, the new parametrization of the CTE loss is accurate within 5\% for
any data point in the spectra used here, while the RMS accuracy for all
spectra used in this study together stays within 1\%. For reference, the
dotted lines in panel (e) in Figure~\ref{f:cti_specplot} depict the Poisson
noise associated with a resolution element (assumed to be 2 pixels along the
dispersion) in spectra of a given signal level.   
{\it This CTE correction formula renders spectroscopic flux calibration to an
  accuracy better than the uncertainty due to Poisson noise per resolution
  element}.

\subsubsection{Imaging Mode} \label{s:ctifunc_i}

CTI values are calculated for every epoch of the external sparse field
test as explained in Sect.~\ref{s:ESFtest}, grouped by ranges of flux and
sky background levels. The ranges of signal levels are chosen after inspecting
histograms of signal levels for each sky level listed in
Table~\ref{t:ESFobs}, aiming for similar numbers of targets in each
group. Table~\ref{t:imagCTItab} lists the derived CTI values for each group
and epoch. 

\paragraph{Time Dependence: Consistency With Spectroscopic Mode} 

\begin{figure}
%\vspace*{12cm}
\centerline{%\psfig{file=f15.eps,width=15cm}
\includegraphics[width=\textwidth]{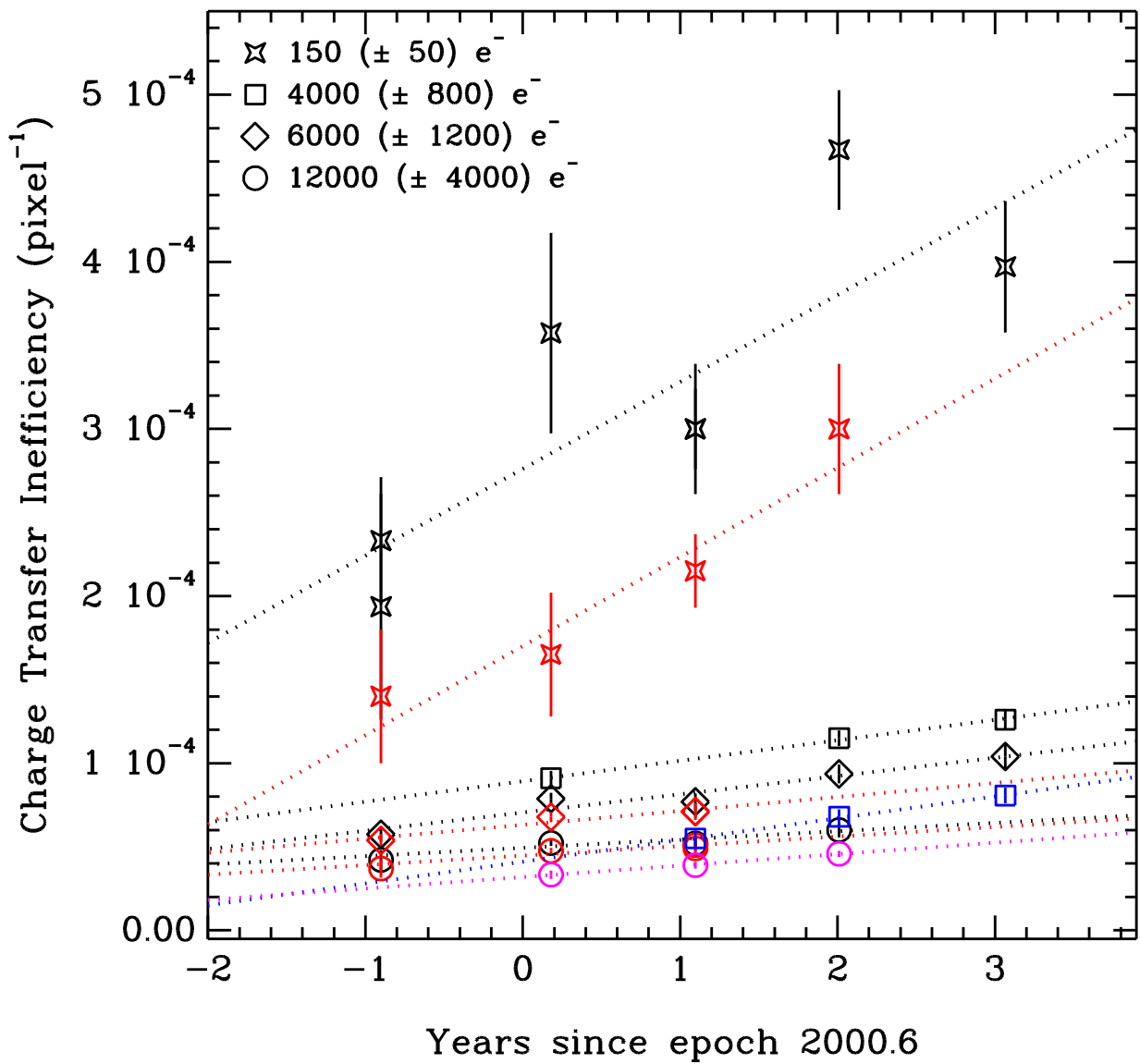}
}
\caption{Imaging CTI as a function of time, derived from the external sparse field
  test. Both the data and the corresponding linear fits are plotted (the
  latter in dotted lines). Symbols associated with individual signal
  levels are indicated in the legend (at the top left). Data with different sky
  background have symbols with different colors: Black for sky = 3 e$^-$
  pixel$^{-1}$, red for sky = 6 e$^-$ pixel$^{-1}$, blue for sky = 10 e$^-$
  pixel$^{-1}$, and magenta for sky = 30 e$^-$
  pixel$^{-1}$. \label{f:cti_vs_time_i}}       
\end{figure}

Before embarking on a comprehensive procedure to fit the imaging CTI values
as a function of signal and background levels, we derive the time
dependence of the CTE loss from these data and check for consistency with the time
constant derived before in Sect.~\ref{s:timedep} for spectroscopic mode. To
this end, we select groups of signal and sky levels that were measured
consistently in at least three epochs and whose CTI uncertainties were
smaller than the CTI values themselves. The CTI values of these groups are
plotted in Figure~\ref{f:cti_vs_time_i} as a function of time, along with
least-squares fits of CTI vs.\ time (separately to each group). After 
weighting the slopes of the individual fits by their inverse variances, the
weighted average time constant is 0.214\,$\pm$\,0.047, i.e.,
consistent with the 
0.205\,$\pm$\,0.006 found in Sect.~\ref{s:timedep}. We adopt the latter in the
following, given the much longer time baseline and the more homogeneous and
well-determined data used to derive it.  

\paragraph{Dependence on Signal and Background Level} 

The functional form to characterize the CTI of the STIS CCD in imaging
mode was determined in a way similar to that of the spectroscopic mode. 
We first attempted to fit the imaging CTI values with the function in
Eq.~\ref{eq:cti_B_G} used to fit the spectroscopic CTI values, but found
that the dependence on background level required more terms in this
case. After extensive experimentation, the following functional 
form produced the best fit to the imaging data:  
\begin{equation}
\mbox{CTI}_{\it IM} =  a\, {\rm e}^{-b\;{\rm lcts}} \times (c\,(t-t_0) + 1) 
 \times \left[ d \exp\,(-e\;{\rm lbck}) + 
 (1-d) \exp\left(-f \left( \frac{{\rm bck}}{{\rm counts}} \right)^{\it g} \right) 
   \right] 
\label{eq:imagCTIform}
\end{equation}
where `counts' stands for the net counts within the measurement aperture,
and 
%\vspace*{-4.ex}
\begin{eqnarray}
t - t_0 & \equiv & ({\rm MJD} - 51765)/365.25 \nonumber \\
{\rm lcts} & \equiv & \ln ({\rm counts}) - 8.5 \nonumber \\
{\rm bck} & \equiv & {\rm max} (0.0,\,{\rm sky}) \nonumber \\
{\rm lbck} & \equiv & \ln ({\rm sqrt} ({{\rm bck}^2 + 1})) - 2 \nonumber
\end{eqnarray}
The constants 51765, 8.5, and 2 were roughly the averages of the corresponding
parameters in the data, and were included to provide numerical stability as
well as to produce independent coefficients ($a$ through $g$). The purpose of 
the `bck' and `lbck' parameters is to avoid logarithms of negative
values. (The parameter $c$ was fixed to 0.205 as discussed above.) 
The fitting procedure was the same as that used in Sect.~\ref{s:spec_cti}
above. The resulting best-fit values of the coefficients in Eq.\
\ref{eq:imagCTIform} are listed in Table~\ref{t:imagcticoeff}. 

\begin{figure}
\centerline{%\psfig{file=f16.eps,width=1.0\textwidth}
\includegraphics[width=\textwidth]{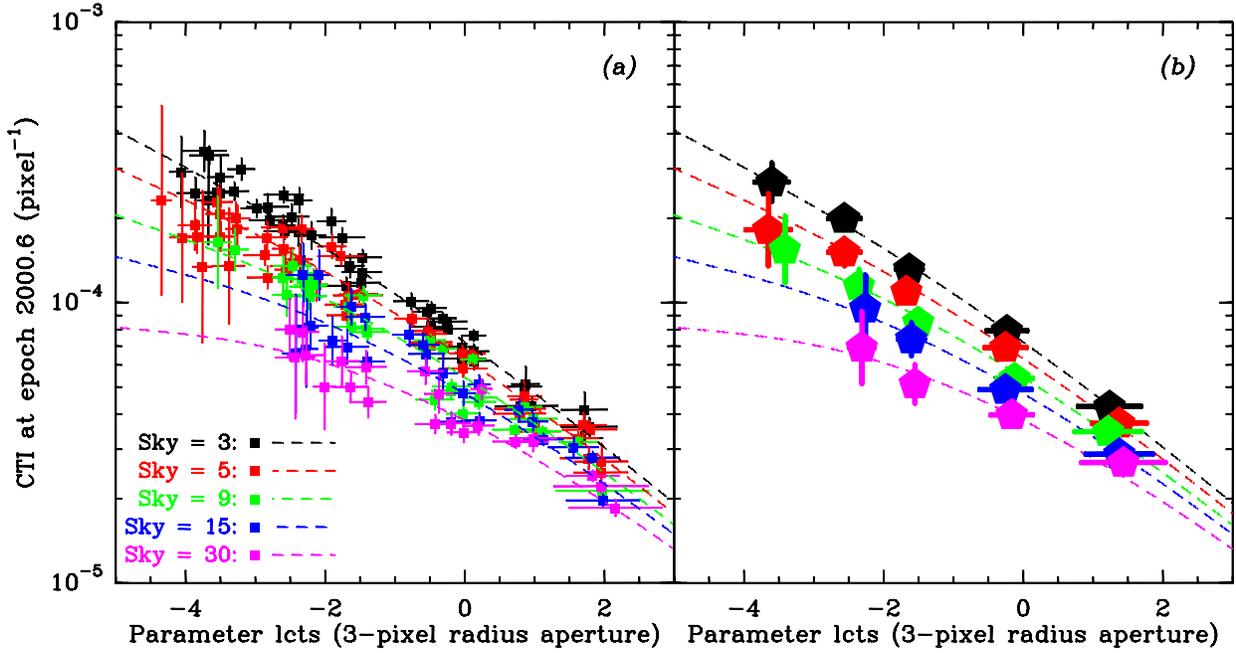}
}
\caption{CTI at epoch 2000.6 as a function of parameter `lcts' (as defined in
  Sect.~\ref{s:ctifunc_i}) for imaging data derived from the external sparse field
  test. Panel (a) shows all individual data points listed in
  Table~\ref{t:imagCTItab}, while panel (b) shows {\it weighted
  mean\/} data points (averaged over 5 ranges of `lcts', namely $-3
  <$~lcts, $-3 \le \mbox{lcts} < -2$, $-2 \le \mbox{lcts} < -1$, $-1
  \le \mbox{lcts} < 0.5$, and $\mbox{lcts} \ge 0.5$, separately
  for each sky level). The drawn lines in both panels depict the fit
  of Eq.~\ref{eq:imagCTIform} to the data. Symbol and line colors
  associated with individual sky levels (in e$^-$ pixel$^{-1}$) are indicated
  in the legend in panel (a) (at the bottom left). \label{f:ESF_cti_vs_lcts}}      
\end{figure}

The quality of this parametrization of the CTE correction is depicted in
Figure~\ref{f:ESF_cti_vs_lcts}. Quantitatively, it yields a correction that
is accurate within 4\,$\sigma$ of {\it any\/} single data point in
Table~\ref{t:imagCTItab}, while the overall RMS uncertainty is 3.0\%, which is
equivalent to 0.2\,$\sigma$ of the ``mean'' data points plotted in 
panel (b) of Figure~\ref{f:ESF_cti_vs_lcts}. To put this in perspective, an 
observation of a typical faint star in September 2002 with an
extracted signal of 100 e$^-$ and a background of 6 e$^-$ per pixel at
the center of the CCD underwent a CTE loss of $\sim$\,15\% (cf.\
Figure~\ref{f:cti_vs_time_i}).  The parametrization in
Eq.~\ref{eq:imagCTIform} corrects this loss to an RMS photometric
accuracy of only 0.5\%, {\it much lower than that due to
  Poisson noise for that object}.   

\paragraph{Centroid Shifts in Imaging Mode}

Centroid shifts were also measured for the external sparse field imaging
data, defined (as before) as half the difference between the centroids of the
target stars as measured by the two different amplifiers. 

\begin{figure}
\centerline{\includegraphics*[width=15cm]{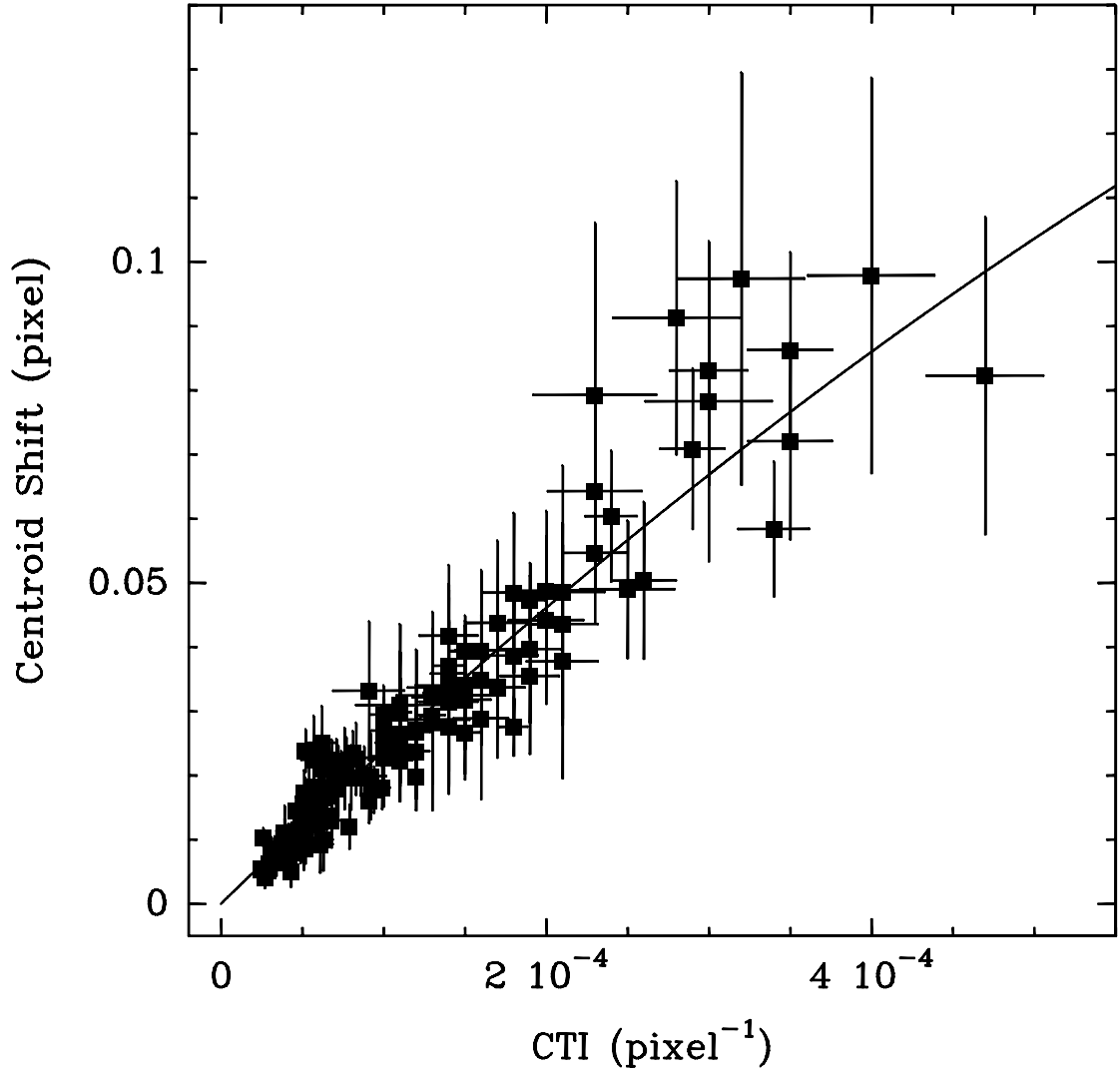}}
\caption{The centroid shift at the central row of the CCD as a function of
  CTI for the imaging data taken during the external sparse field tests. A
  weighted least-squares fit to the data (cf.\ Eq.~\ref{eq:shift_from_cti_i})
  is shown by the solid line.}   
\label{f:cti_vs_shift_i}
\end{figure}

The derived centroid shifts are plotted against measured CTI in
Figure~\ref{f:cti_vs_shift_i}, which shows a strong relation between
the two parameters, similar to the case for the spectroscopic data described
in Sect.~\ref{s:timedep}. 
The solid line depicts a weighted least-square fit to this relation for the imaging
data, 
\begin{equation}
\mbox{Centroid Shift [pixels]} = 0.025 \, \left(
  \frac{\mbox{CTI}}{10^{-4}} \right) \,
 - \, 0.78\: 10^{-3}\, \left( \frac{\mbox{CTI}}{10^{-4}} \right)^2 
\label{eq:shift_from_cti_i}
\end{equation}
which has an RMS error of 0.005 pixels\footnote{Again, positive centroid
shifts in this equation indicate shifts toward {\it smaller\/} Y coordinate
values.}. We emphasize that centroid shifts due to CTE effects are rarely
taken into account \citep[a literature search using the  NASA Astrophysics
Data System\footnote{http://www.adsabs.harvard.edu} came up with only one
study:][]{piat+05}, whereas they are relevant for science programs that aim
to reach the highest possible positional accuracy such as proper motion
measurements for which accuracy of order 0.01 pixel is often required
\citep*[e.g.,][]{mcna+03,kosl+06}. This is especially important if
one combines measurements done with two different instruments (or with one
instrument but at significantly different locations on the CCD) to derive
proper motions.  

A comparison of Eq.~\ref{eq:shift_from_cti_i} with Eq.~\ref{eq:shift_from_cti} 
(or, equivalently, Figure~\ref{f:cti_vs_shift_i} with
Figure~\ref{f:cti_vs_shift}) shows that the imaging data incur a roughly 
3 times smaller centroid shift than spectroscopic data for a given CTI value. 
We interpret this difference in the sense that the measured centroid shifts
are most heavily influenced by the centroid shift incurred by the {\it peak
  pixel\/} of the PSF (since it is by far the brightest pixel within 
the measurement aperture), and higher signal levels cause lower centroid
shifts. To give an illustrative estimate of this effect: A comparison of
Eq.~\ref{eq:imagCTIform} with Eq.~\ref{eq:fin_specCTI} shows that for a
given CTI value and background level, the signal level in imaging mode is a factor
$\sim$\,7 higher than in spectroscopic mode. Approximating the imaging and
spectroscopic PSFs by Gaussians with a FWHM of 1.8 pixels, this
translates to a ratio in peak intensity 
\[ \frac{{\it Peak}_{\it IMA}}{{\it Peak}_{\it SPEC}} \; = \; \frac{7}{\sqrt{\pi/(4
  \ln 2)}\;{\rm FWHM}_{\rm PSF}} \; \sim \; 3.6. 
\]
According to
Eq.~\ref{eq:fin_specCTI}, a 3.6 times higher signal level causes a
$\sim$\,3.0 times lower CTI value, and hence a 2.9 times lower centroid shift (cf.\
Eq.~\ref{eq:shift_from_cti}), which is consistent with what we find in practice. 

\paragraph{The Effect of Aperture Size for Imaging Mode}

\begin{figure}
%\vspace*{12cm}
\centerline{%\psfig{file=f18.eps,width=1.0\textwidth,angle=-90.}
\includegraphics[angle=-90,width=\textwidth]{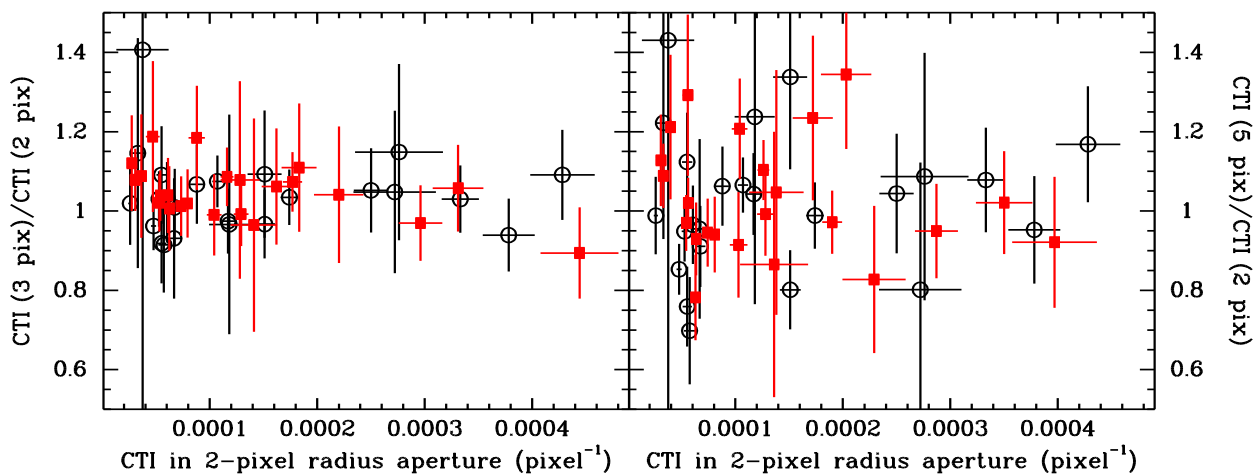}
}
\caption{The ratio between imaging CTI values measured in different
  aperture sizes. {\it Left panel}: 3-pixel radius vs.\ 2-pixel
  radius. {\it Right panel}: 5-pixel radius vs.\ 2-pixel radius. Open
  circles indicate measurements from the Sep 2002 epoch, while filled
  squares indicate measurements from the Sep 2003 epoch. 
  The size of the measurement aperture does not have any significant 
  impact on the CTI in imaging mode.\label{f:ESF_apereffect}}      
\end{figure}

Since the amplitude of the centroid shift increases with CTE loss as discussed
above, one obvious question is whether this effect has an impact on
small-aperture photometry, which is often used in imaging programs of compact
sources in order to achieve an increased signal-to-noise ratio relative to
the use of larger apertures. We checked for this by evaluating 
the dependence of the measured CTE loss on aperture size using the imaging data
of the external sparse field test. Photometry measurements were made
through three aperture sizes (radii of 2, 3,  and 5 pixels). The result is
illustrated in Figure~\ref{f:ESF_apereffect}. There is no significant
difference in CTI value among the three apertures used: The formal ratio between
CTI values measured through 3-pixel radius and through 2-pixel radius
apertures for the data shown in  Figure~\ref{f:ESF_apereffect} is 0.998
$\pm$ 0.015 (mean error of the mean). Likewise, the ratio between CTI values
measured through 5-pixel radius and through 2-pixel radius apertures is 1.031
$\pm$ 0.020. We interpret this result in the sense that the centroid shift to
the PSF caused by CTE effects does not cause actual loss of signal as long as
one measures the centroids of the targets on the image itself with an
adequately large centering box width\footnote{We used the   default size of
  the centering box within {\sc daophot-ii} ({\sc cbox} = 5 pixels).}.  

\subsection{Implementation of CTE Correction} \label{s:implement}

As we have adopted the definition of CTE as the fraction of charge
transferred per pixel transfer, the correction for CTI ($\equiv 1 -
\mbox{CTE}$) to be applied to an extracted STIS data point {\it OBJ\/} (be it
a flux element of a STIS CCD spectrum\footnote{Note however that since
  December 16, 2003, the pipeline flux calibration of STIS CCD spectra takes
  correction for CTI already into account for full-frame datasets
  \citep{gouboh06}.} or an  entry of a star in a photometry table from STIS
CCD imaging) is  
\begin{equation}
\mbox{\it OBJ}_{CORR} = \frac{\mbox{\it OBJ}}{(1 - \mbox{\it
    CTI}_{\it OBJ})\,^{1024 - \mbox{\it y}\,*\mbox{\it ybin}}} 
\label{eq:implement}
\end{equation}
where {\it y\/} is the Y coordinate of the centroid of the target, {\it
  ybin\/} is the on-chip binning factor, and {\it CTI}$_{\it OBJ}$ is the result
of Eq.~\ref{eq:fin_specCTI} [for spectroscopic data] or
Eq.~\ref{eq:imagCTIform} [for imaging data] for that data point.

\section{Concluding Remarks}   \label{s:concl}

Using a variety of methods, a set of formulae to correct observations of
point sources with the STIS CCD for CTE-related loss of signal has been
derived and presented. For data taken in imaging mode, the CTE loss is
parametrized in terms of the location of the source on the CCD, the source
signal level within the measurement aperture, the background level, and the
time of observation. For spectroscopic data, one extra parameter is needed to
provide an adequate calibration of the CTE loss, namely the signal in the
point spread function located between the signal extraction box and the
read-out amplifier. The effect of the latter parameter is significant at
wavelengths redward of 8000 \AA\ where the PSF of the STIS CCD exhibits a
extended ``halo'', i.e., for spectra taken using the G750L or G750M gratings
of STIS. The algorithms presented here correct flux calibration inaccuracies
due to CTE losses as large as 30\% to within $\sim$\,1.5\% RMS throughout the
wavelength range covered by the STIS CCD observing modes. 

Using bi-directional CCD readouts, centroid shifts incurred due to CTE loss
are also derived. Centroid shifts are up to $\sim$\,10$^{-1}$ pixel for
imaging photometry of faint point sources, and a factor of $\sim$\,3 higher
for spectroscopy mode. A strong correlation is found between the signal loss
and the centroid shift (both for imaging and spectroscopic modes), thus
enabling one to correct for both effects of imperfect CTE to STIS CCD
observations. While centroid shifts due to CTE effects are
rarely taken into account in the literature, they are relevant for
science programs that aim to reach the highest possible positional accuracy
such as proper motion measurements for which accuracy of order 0.01 pixel is
often required. 

Though we have demonstrated that the correction formulae work well, it is
of course preferable (when practical) to minimize the magnitude of the CTE 
effects on an observing program at the data acquisition stage. To this
end, STIS operations support acquisition of targets for CCD spectroscopic
modes at the so-called `E1 pseudo-apertures', which are located $\sim$\,125
rows from the readout amplifier rather than at field center. The
appropriate instrumental calibrations have been developed for these aperture positions. 
At the STIS plate scale of 0.05 arcseconds per pixel 
along the slit, this position gives 6 arcseconds of sky coverage to the  
near edge of the CCD; targets for which this angular coverage is sufficient
can thus be observed with four times lower charge loss by selecting this
observing position rather than the center of the CCD. Corrections for CTE
loss and the use of the `E1 pseudo-apertures' will be of even greater
importance in the future if STIS is successfully repaired as planned  in
the hoped-for HST Servicing Mission 4, as the CCD usage will be extended
for a number of additional years in the orbital radiation environment.  

\paragraph*{Acknowledgments}

We are grateful to the anonymous referee for a quick and thorough review, and
for making constructive suggestions that improved the paper. 
We acknowledge the valuable contributions of Tom Brown,
Ron Gil\-li\-land, and Mike Potter to STIS CTE analysis in the past.

\clearpage


\clearpage

\clearpage

\clearpage 

\clearpage 

\clearpage 

\clearpage 

\clearpage 

\clearpage 

\clearpage 

\clearpage 

\clearpage

\clearpage 

\clearpage 

\clearpage

\clearpage

\clearpage

\clearpage 

\clearpage 

%% If you are not including electronic art with your submission, you may
%% mark up your captions using the \figcaption command. See the 
%% No more than seven \figcaption commands are allowed per page, 
%% so if you have more than seven captions, insert a \clearpage 

%% Tables should be submitted one per page, so put a \clearpage before

%% Two options are available to the author for producing tables:  the
%% deluxetable environment provided by the AASTeX package or the LaTeX
%% table environment.  Use of deluxetable is preferred.
%%

%% Three table samples follow, two marked up in the deluxetable environment,
%% one marked up as a LaTeX table.

%% In this first example, note that the \tabletypesize{}
%% command has been used to reduce the font size of the table.
%% Note also that the \label command needs to be placed 

\clearpage

%
% Table of STIS Observations
%
\begin{deluxetable}{cccccr}
\tabletypesize{\small}
\tablecaption{STIS observations for ``internal'' sparse field
  tests\label{t:ISFobs}} 
\tablewidth{0pt}
\tablecolumns{6}
\tablehead{
\colhead{Optical} & \colhead{$\lambda_{\rm cen}$} & \colhead{Slit} & 
\colhead{Exp.\ Time} & 
\colhead{CCD Gain} & \colhead{Signal}  \\
%\colhead{$\Theta$\tablenotemark{b}}
\colhead{Element} & \colhead{({\AA})} & \colhead{Name} & \colhead{(s)}  &
\colhead{(e$^-$/DN)}  & \colhead{Levels\tablenotemark{a}}   
}
% startJD = 2450999.779, startMJD = 50999.279
\startdata
G430M  & 5471 & 0.05x31NDA & 0.3 & 1, 4 & 60 \\
G430M  & 5471 & 0.05x31NDA & 0.6 & 1, 4 & 130 \\
G430M  & 5471 & 0.05x31NDA & 0.9 & 1, 4 & 195 \\
G430M  & 5471 & 0.05x31NDA & 2.3 & 1, 4 & 500 \\
Mirror & ---\tablenotemark{b} & 0.05x31NDA & 0.3 & 1, 4 & 3400 \\
Mirror & ---\tablenotemark{b} & 0.05x31NDB & 0.3 & 1, 4 & 9800 \\ [0.5ex] 
\enddata

%% Text for table notes should follow after the \enddata but before
%% the \end{deluxetable}. Make sure there is at least one \tablenotemark
%% in the table for each \tablenotetext.

\tablenotetext{a}{Unit is e$^-$ per column {\it in the absence of CTE losses}.}
\tablenotetext{b}{Using ``Clear'' filter (i.e., no filter; aperture name 50CCD)}

\tablecomments{All exposures were taken in two read modes: Once read out using
  the default amplifier ``D'', and once using amplifier ``B'', on the opposite
  side of the parallel register.}

\end{deluxetable}

\begin{deluxetable}{ccccc}
\tabletypesize{\small}
\tablecaption{Observing blocks used for the internal sparse field test. Each
  block extended over a time period of one to a few days. Representative
  values for the Modified Julian Date (MJD) and the civil date are
  shown.\label{t:obsblocks}}  
\tablewidth{0pt}
\tablecolumns{6}
\tablehead{
\colhead{Gain} & \colhead{Block} & \colhead{Program} & \colhead{MJD} & 
\colhead{Date (UT)}
}
\startdata
1    & 1     & 8414    & 51428 & Sep 07, 1999 \\
1    & 2     & 8414    & 51650 & Apr 16, 2000 \\
1    & 3     & 8851    & 51845 & Oct 28, 2000 \\
1    & 4     & 8910    & 52210 & Oct 28, 2001 \\
1    & 5     & 9620    & 52567 & Oct 20, 2002 \\
1    & 6     & 9620    & 52896 & Sep 14, 2003 \\
\tableline
4    & 1     & 8414    & 51783 & Aug 26, 2000 \\
4    & 2     & 8851    & 52004 & Apr 04, 2001 \\
4    & 3     & 8910    & 52399 & May 05, 2002 \\
4    & 4     & 9620    & 52734 & Apr 05, 2003 \\
4    & 5     & 10026   & 53099 & Apr 04, 2004 \\
\enddata
\end{deluxetable}

\begin{deluxetable}{cccrcl}
\tabletypesize{\small}
\tablecaption{STIS observations\tablenotemark{a}\ for ``external'' sparse imaging field
  test\label{t:ESFobs}} 
\tablewidth{0pt}
\tablecolumns{6}
\tablehead{
\colhead{Observing} & \colhead{On-orbit Time} &
\colhead{Exp.\ Time} & 
\colhead{$N_{\rm exp}$} & \colhead{CCD Gain} & \colhead{Background}  \\
%\colhead{$\Theta$\tablenotemark{b}}
\colhead{Date} & \colhead{(yr)} & \colhead{(s)} & \colhead{\ }  &
\colhead{(e$^-$/DN)}  & \colhead{Levels\tablenotemark{b}}   
}
% startJD = 2450999.779, startMJD = 50999.279
\startdata
Sep 15, 1999 & 2.54 &         20 & 24 & 1 & 3.1, 5.1 \\
             &      & \llap{1}00 &  9 & 1 & 14 \\
Oct 13, 2000 & 3.62 &         20 & 40 & 1 & 3.3, 5.9, 7.9 \\
             &      & \llap{1}00 & 12 & 1 & 16, 27 \\
Sep 13, 2001 & 4.54 &         20 & 40 & 1 & 3.0, 5.8, 7.9 \\
             &      & \llap{1}00 & 12 & 1 & 11, 37 \\
Sep 13, 2002 & 5.45 &         20 & 40 & 1 & 2.9, 5.3 \\
             &      & \llap{1}00 & 12 & 1 & 11, 33 \\
Sep 13, 2003 & 6.51 &         20 & 40 & 1 & 3.4  \\
             &      & \llap{1}00 & 12 & 1 & 11, 16 \\
\enddata

%% Text for table notes should follow after the \enddata but before
%% the \end{deluxetable}. Make sure there is at least one \tablenotemark
%% in the table for each \tablenotetext.
\tablenotetext{a}{The imaging field in NGC 6752 was centered on 
  $\alpha$ = 19$^{\rm d}$11$\!^{\rm m}$05\fs96,
  $\delta$ = $-$60\degr01$'$\,06\farcs7 in J2000 equinox. Observations were
  conducted using the   ``Clear'' filter (i.e., no filter; aperture name {\tt
  50CCD}).} 
\tablenotetext{b}{Centroids of histogram of sky pixel values windowed to
  $\pm 4 \sigma$ of the mean. Unit is e$^-$ per pixel.}

\tablecomments{All exposures were taken in two read modes: Once read out using
  the default amplifier ``D'', and once using amplifier ``B'', on the opposite
  side of the parallel register. Celestial coordinates are given in equinox J2000.}

\end{deluxetable}

\begin{table}[p]
\caption{CTI time constant $\alpha$ as measured from the internal
  sparse field test data taken in gain\,=\,1, as a function of signal level.}
                                %The last row lists our adopted value in
                                %boldface font.} 
\label{t:alphaval}
\bigskip 
\centerline{\begin{tabular*}{5cm}{@{\extracolsep{\fill}}rcc@{}}
\tableline \tableline
 signal & $\alpha$    &  $\sigma_\alpha$ \\ 
 (DN)   & (yr$^{-1}$) &  (yr$^{-1}$)     \\ [0.5ex]
\tableline
\multicolumn{3}{c}{~} \\ [-2.8ex]
  60 & 0.216 & 0.009 \\
 130 & 0.192 & 0.013 \\
 195 & 0.188 & 0.021 \\
 500 & 0.202 & 0.016 \\
3450 & 0.218 & 0.038 \\
9850 & 0.170 & 0.052 \\
\tableline
\end{tabular*}}
\end{table}
%
% FYI: Formal weighted mean is 0.205 +- 0.006 (mean error of mean)
%

\begin{table}[p]
\small 
\caption{List of flux standard star spectra used to characterize the CTI
  as function of signal and background level. All
  intensities are in e$^-$ per pixel per (sub-)exposure.} 
\label{t:spectra}
\begin{center}
\begin{tabular*}{\textwidth}{@{\extracolsep{\fill}}lllcc@{~~~}c@{}}
\multicolumn{3}{c}{~} \\ [-4ex]
\hline \hline
\multicolumn{3}{c}{~} \\ [-2.2ex]
 Rootname & Grating & Flux Standard & Observing & Background & 
 \multicolumn{1}{c}{Range in} \\ 
          &         &               & Date      & level      & 
 Signal Levels \\ [0.5ex]
\hline
\multicolumn{3}{c}{~} \\ [-2.2ex]
\tt o6ig10010 & G230LB & G191\,--\,B2B & 2002--04--22 & 0.4 & \llap{10}00 -- 5000 \\
\tt o6ig01060 & G750L  & GD\,71 & 2001--12--07 & 1.7 & \llap{3}00 -- 1600\rlap{0}  \\
\tt o6il01020 & G230LB & LDS\,749B  & 2001--07--13 & 1.9 & \llap{1}00 -- 1800 \\
\tt o8u2200b0 & G430L  & AGK+81D266 & 2003--11--03 & 0.5 & 
 \llap{30}00 -- 9200 \\ 
\tt o8v101030 & G750L  & WD\,1657+343 & 2004--04--07 & 2.5 & ~\,30 -- 750~~~ \\
\tt o8v2040e0 & G230LB & GD\,71    & 2003--10--24 & 0.3 & \llap{1}20 -- 730~\, \\
\tt o8v204030 & G230LB & GD\,71    & 2003--10--24 & 0.1 &  ~\,20 -- 170~~~ \\ [0.5ex]
\hline
\multicolumn{3}{c}{~} \\ [-5ex]
\end{tabular*}
\end{center}
\end{table}

\clearpage

\begin{table}
\caption{Best-fit Values of Coefficients in Functional Form of Spectroscopic CTI
  (Eq.~\ref{eq:fin_specCTI}).}  
\label{t:speccticoeff}
\centerline{
\small
\begin{tabular}{@{}cc@{~~~~~~~~}l@{}}
\multicolumn{3}{c}{~} \\ [-2.5ex]   
\tableline \tableline
\multicolumn{3}{c}{~} \\ [-1.8ex]                                                
Coefficient & Value & Description \\ [0.5ex] 
\tableline 
\multicolumn{3}{c}{~} \\ [-2.ex]   
$\alpha$   & 0.056 $\pm$ 0.001 & CTI normalization \\
$\beta$    & 0.82 $\pm$ 0.01 & Gross count level dependence \\
$\gamma$   & 0.205 $\pm$ 0.006 & Time dependence of CTI \\
$\delta$   & 3.00 $\pm$ 0.05 & Multiplicative factor for `background'/gross count ratio \\
$\epsilon$ & 1.30 $\pm$ 0.10 & Multiplicative factor for halo light fraction \\
$\zeta$    & 0.18 $\pm$ 0.01 & Power of `background'/gross count ratio \\ 
$\eta$     & 0.06 $\pm$ 0.01 & Minimum value of halo light
 fraction above spectrum \\ [0.5ex]
\tableline 
\end{tabular}
}
\end{table}

\clearpage 

\begin{table}
\caption{Measured CTI values and centroid shifts for external sparse imaging field
  test data.}  
\label{t:imagCTItab}
\centerline{
\small
\begin{tabular}{@{}ccr@{\,$\pm$\,}lcc@{}}
\multicolumn{3}{c}{~} \\ [-2.ex]   
\tableline \tableline
\multicolumn{3}{c}{~} \\ [-2.5ex]                                                
MJD & Sky Level & \multicolumn{2}{c}{Signal Level} &
 \multicolumn{1}{c}{CTI} & Centroid Shift  \\
    & (e$^-$/pixel) & \multicolumn{2}{c}{(e$^-$)} & (pixel$^{-1}$) &
 (pixel) \\ [0.5ex] 
\tableline 
\multicolumn{3}{c}{~} \\ [-2.ex]   
 51436 &  3.1 &   149 &   27 & (2.3 $\pm$ 0.4) 10$^{-4}$ & 0.0793 $\pm$ 0.0268 \\
       &  3.1 &   293 &   87 & (1.8 $\pm$ 0.2) 10$^{-4}$ & 0.0485 $\pm$ 0.0124 \\
       &  3.1 &   460 &   82 & (1.9 $\pm$ 0.2) 10$^{-4}$ & 0.0397 $\pm$ 0.0114 \\
       &  3.1 &   852 &  276 & (1.4 $\pm$ 0.1) 10$^{-4}$ & 0.0359 $\pm$ 0.0084 \\
       &  3.1 &  2293 &  567 & (8.3 $\pm$ 0.5) 10$^{-5}$ & 0.0226 $\pm$ 0.0056 \\
       &  3.1 &  2888 &  593 & (7.6 $\pm$ 0.5) 10$^{-5}$ & 0.0211 $\pm$ 0.0063 \\
       &  3.1 &  4708 &  913 & (5.7 $\pm$ 0.4) 10$^{-5}$ & 0.0183 $\pm$ 0.0070 \\
       &  5.1 &   290 &   85 & (1.4 $\pm$ 0.2) 10$^{-4}$ & 0.0418 $\pm$ 0.0110 \\
       &  5.1 &   476 &   84 & (1.5 $\pm$ 0.2) 10$^{-4}$ & 0.0318 $\pm$ 0.0115 \\
       &  5.1 &   836 &  274 & (1.2 $\pm$ 0.1) 10$^{-4}$ & 0.0268 $\pm$ 0.0073 \\
       &  5.1 &  2320 &  569 & (7.2 $\pm$ 0.4) 10$^{-5}$ & 0.0192 $\pm$ 0.0055 \\
       &  5.1 &  2902 &  587 & (6.5 $\pm$ 0.4) 10$^{-5}$ & 0.0182 $\pm$ 0.0053 \\
       &  5.1 &  4765 &  876 & (5.4 $\pm$ 0.4) 10$^{-5}$ & 0.0158 $\pm$ 0.0060 \\
       & 14.1 &   915 &  279 & (5.7 $\pm$ 0.8) 10$^{-5}$ & 0.0225 $\pm$ 0.0068 \\ 
       & 14.1 &  2231 &  532 & (6.3 $\pm$ 0.6) 10$^{-5}$ & 0.0100 $\pm$ 0.0048 \\ 
       & 14.1 &  4833 &  838 & (3.9 $\pm$ 0.4) 10$^{-5}$ & 0.0111 $\pm$ 0.0042 \\ 
       & 14.1 & 10721 & 2157 & (3.4 $\pm$ 0.3) 10$^{-5}$ & 0.0086 $\pm$ 0.0028 \\ 
       & 14.1 & 23397 & 8367 & (2.5 $\pm$ 0.2) 10$^{-5}$ & 0.0055 $\pm$ 0.0019 \\ [0.5ex]
 51831 &  3.3 &   413 &  101 & (2.1 $\pm$ 0.2) 10$^{-4}$ & 0.0436 $\pm$ 0.0157 \\ 
       &  3.3 &   951 &  193 & (1.4 $\pm$ 0.1) 10$^{-4}$ & 0.0276 $\pm$ 0.0069 \\ 
       &  3.3 &  1141 &  308 & (1.5 $\pm$ 0.1) 10$^{-4}$ & 0.0267 $\pm$ 0.0073 \\ 
       &  3.3 &  3055 &  659 & (9.9 $\pm$ 0.5) 10$^{-5}$ & 0.0181 $\pm$ 0.0033 \\ 
       &  3.3 &  3586 &  689 & (9.1 $\pm$ 0.4) 10$^{-5}$ & 0.0161 $\pm$ 0.0035 \\ 
       &  3.3 &  5600 &  957 & (7.9 $\pm$ 0.4) 10$^{-5}$ & 0.0120 $\pm$ 0.0034 \\ 
       &  3.3 & 11594 & 2812 & (5.2 $\pm$ 0.2) 10$^{-5}$ & 0.0086 $\pm$ 0.0023 \\ 
       &  3.3 & 26075 & 8100 & (3.4 $\pm$ 0.2) 10$^{-5}$ & 0.0072 $\pm$ 0.0028 \\ 
       &  5.9 &   407 &  119 & (1.4 $\pm$ 0.2) 10$^{-4}$ & 0.0340 $\pm$ 0.0158 \\ 
       &  5.9 &   931 &  195 & (1.1 $\pm$ 0.1) 10$^{-4}$ & 0.0309 $\pm$ 0.0084 \\ 
       &  5.9 &  1156 &  315 & (1.1 $\pm$ 0.1) 10$^{-4}$ & 0.0265 $\pm$ 0.0076 \\ 
       &  5.9 &  3049 &  675 & (8.0 $\pm$ 0.4) 10$^{-5}$ & 0.0197 $\pm$ 0.0050 \\
       &  5.9 &  3661 &  689 & (7.1 $\pm$ 0.4) 10$^{-5}$ & 0.0179 $\pm$ 0.0053 \\ 
       &  5.9 &  5585 &  925 & (6.8 $\pm$ 0.3) 10$^{-5}$ & 0.0169 $\pm$ 0.0052 \\ 
       &  5.9 & 11666 & 2821 & (4.8 $\pm$ 0.2) 10$^{-5}$ & 0.0116 $\pm$ 0.0048 \\ [0.5ex]
\tableline
\end{tabular}
}
\end{table}
\addtocounter{table}{-1}

\clearpage

\begin{table}
\caption{(continued)}  
\centerline{
\small
\begin{tabular}{@{}ccr@{\,$\pm$\,}lcc@{}}
\multicolumn{3}{c}{~} \\ [-2.ex]   
\tableline \tableline
\multicolumn{3}{c}{~} \\ [-2.5ex]                                                
MJD & Sky Level & \multicolumn{2}{c}{Signal Level} &
 \multicolumn{1}{c}{CTI} & Centroid Shift  \\
    & (e$^-$/pixel) & \multicolumn{2}{c}{(e$^-$)} & (pixel$^{-1}$) &
 (pixel) \\ [0.5ex] 
\tableline 
\multicolumn{3}{c}{~} \\ [-2.ex]   
 51831 &  7.9 &   417 &  113 & (1.4 $\pm$ 0.2) 10$^{-4}$ & 0.0315 $\pm$ 0.0128 \\ 
       &  7.9 &   930 &  183 & (1.0 $\pm$ 0.1) 10$^{-4}$ & 0.0228 $\pm$ 0.0076 \\ 
       &  7.9 &  1153 &  303 & (1.1 $\pm$ 0.1) 10$^{-4}$ & 0.0224 $\pm$ 0.0064 \\ 
       &  7.9 &  3007 &  661 & (7.6 $\pm$ 0.4) 10$^{-5}$ & 0.0199 $\pm$ 0.0034 \\ 
       &  7.9 &  3622 &  691 & (7.1 $\pm$ 0.4) 10$^{-5}$ & 0.0180 $\pm$ 0.0032 \\ 
       &  7.9 &  5546 &  945 & (6.5 $\pm$ 0.3) 10$^{-5}$ & 0.0138 $\pm$ 0.0023 \\ 
       &  7.9 & 11656 & 2819 & (4.4 $\pm$ 0.2) 10$^{-5}$ & 0.0104 $\pm$ 0.0019 \\ 
       &  7.9 & 25093 & 6925 & (3.3 $\pm$ 0.1) 10$^{-5}$ & 0.0067 $\pm$ 0.0020 \\ 
       & 14.8 &  1188 &  325 & (9.2 $\pm$ 0.1) 10$^{-5}$ & 0.0199 $\pm$ 0.0067 \\ 
       & 14.8 &  2846 &  718 & (6.8 $\pm$ 0.8) 10$^{-5}$ & 0.0130 $\pm$ 0.0042 \\ 
       & 14.8 &  3604 &  742 & (5.8 $\pm$ 0.9) 10$^{-5}$ & 0.0144 $\pm$ 0.0046 \\ 
       & 14.8 &  6047 &  937 & (5.3 $\pm$ 0.4) 10$^{-5}$ & 0.0139 $\pm$ 0.0037 \\ 
       & 14.8 & 13100 & 2780 & (3.9 $\pm$ 0.3) 10$^{-5}$ & 0.0103 $\pm$ 0.0029 \\ 
       & 14.8 & 30780 &11882 & (2.9 $\pm$ 0.1) 10$^{-5}$ & 0.0052 $\pm$ 0.0017 \\ 
       & 29.5 &  1197 &  344 & (6.1 $\pm$ 0.8) 10$^{-5}$ & 0.0169 $\pm$ 0.0058 \\ 
       & 29.5 &  2809 &  714 & (5.9 $\pm$ 0.6) 10$^{-5}$ & 0.0137 $\pm$ 0.0040 \\
       & 29.5 &  3403 &  696 & (4.9 $\pm$ 0.8) 10$^{-5}$ & 0.0124 $\pm$ 0.0049 \\ 
       & 29.5 &  6268 &  917 & (5.1 $\pm$ 0.4) 10$^{-5}$ & 0.0091 $\pm$ 0.0038 \\ 
       & 29.5 & 13111 & 2866 & (3.3 $\pm$ 0.3) 10$^{-5}$ & 0.0087 $\pm$ 0.0018 \\ 
       & 29.5 & 30806 &11776 & (2.5 $\pm$ 0.1) 10$^{-5}$ & 0.0054 $\pm$ 0.0012 \\ [0.5ex]
 52166 &  3.0 &   148 &   30 & (3.0 $\pm$ 0.2) 10$^{-4}$ & 0.0831 $\pm$ 0.0178 \\ 
       &  3.0 &   302 &   88 & (2.4 $\pm$ 0.2) 10$^{-4}$ & 0.0604 $\pm$ 0.0102 \\ 
       &  3.0 &   904 &  253 & (1.4 $\pm$ 0.1) 10$^{-4}$ & 0.0371 $\pm$ 0.0069 \\ 
       &  3.0 &  3031 &  882 & (1.0 $\pm$ 0.1) 10$^{-4}$ & 0.0251 $\pm$ 0.0036 \\ 
       &  3.0 &  4739 & 1598 & (7.7 $\pm$ 0.5) 10$^{-5}$ & 0.0224 $\pm$ 0.0036 \\ 
       &  3.0 & 10969 & 3939 & (5.2 $\pm$ 0.3) 10$^{-5}$ & 0.0147 $\pm$ 0.0049 \\ 
       &  3.0 & 29354 &11702 & (4.4 $\pm$ 0.3) 10$^{-5}$ & 0.0091 $\pm$ 0.0036 \\ 
       &  5.8 &   149 &   30 & (2.1 $\pm$ 0.3) 10$^{-4}$ & 0.0486 $\pm$ 0.0197 \\ 
       &  5.8 &   281 &   84 & (1.8 $\pm$ 0.2) 10$^{-4}$ & 0.0387 $\pm$ 0.0108 \\ 
       &  5.8 &   893 &  254 & (1.2 $\pm$ 0.1) 10$^{-4}$ & 0.0238 $\pm$ 0.0066 \\ 
       &  5.8 &  3046 &  901 & (8.8 $\pm$ 0.4) 10$^{-5}$ & 0.0202 $\pm$ 0.0043 \\ 
       &  5.8 &  4811 & 1612 & (7.1 $\pm$ 0.5) 10$^{-5}$ & 0.0197 $\pm$ 0.0060 \\ [0.5ex]
\tableline
\end{tabular}
}
\end{table}
\addtocounter{table}{-1}

\clearpage

\begin{table}
\caption{(continued)}  
\centerline{
\small
\begin{tabular}{@{}ccr@{\,$\pm$\,}lcc@{}}
\multicolumn{3}{c}{~} \\ [-2.ex]   
\tableline \tableline
\multicolumn{3}{c}{~} \\ [-2.5ex]                                                
MJD & Sky Level & \multicolumn{2}{c}{Signal Level} &
 \multicolumn{1}{c}{CTI} & Centroid Shift  \\
    & (e$^-$/pixel) & \multicolumn{2}{c}{(e$^-$)} & (pixel$^{-1}$) &
 (pixel) \\ [0.5ex] 
\tableline 
\multicolumn{3}{c}{~} \\ [-2.ex]   
 52166 &  7.9 &   149 &   31 & (2.1 $\pm$ 0.2) 10$^{-4}$ & 0.0378 $\pm$ 0.0182 \\ 
       &  7.9 &   292 &   83 & (1.5 $\pm$ 0.1) 10$^{-4}$ & 0.0341 $\pm$ 0.0100 \\ 
       &  7.9 &   912 &  255 & (1.1 $\pm$ 0.1) 10$^{-4}$ & 0.0248 $\pm$ 0.0054 \\ 
       &  7.9 &  3042 &  901 & (8.6 $\pm$ 0.3) 10$^{-5}$ & 0.0197 $\pm$ 0.0029 \\ 
       &  7.9 &  4812 & 1620 & (7.1 $\pm$ 0.4) 10$^{-5}$ & 0.0188 $\pm$ 0.0033 \\ 
       &  7.9 & 10801 & 4001 & (4.9 $\pm$ 0.3) 10$^{-5}$ & 0.0103 $\pm$ 0.0039 \\ 
       &  7.9 & 29533 &11383 & (4.3 $\pm$ 0.3) 10$^{-5}$ & 0.0050 $\pm$ 0.0023 \\ 
       & 11.4 &   385 &   76 & (1.3 $\pm$ 0.2) 10$^{-4}$ & 0.0287 $\pm$ 0.0141 \\ 
       & 11.4 &   977 &  236 & (1.0 $\pm$ 0.1) 10$^{-4}$ & 0.0230 $\pm$ 0.0060 \\
       & 11.4 &  3217 &  867 & (5.5 $\pm$ 0.5) 10$^{-5}$ & 0.0130 $\pm$ 0.0032 \\ 
       & 11.4 &  4818 & 1126 & (4.9 $\pm$ 0.3) 10$^{-5}$ & 0.0098 $\pm$ 0.0032 \\ 
       & 11.4 & 10145 & 3215 & (4.3 $\pm$ 0.2) 10$^{-5}$ & 0.0085 $\pm$ 0.0020 \\
       & 11.4 & 35503 &23516 & (2.6 $\pm$ 0.2) 10$^{-5}$ & 0.0054 $\pm$ 0.0019 \\ 
       & 35.2 &   960 &  235 & (6.1 $\pm$ 0.8) 10$^{-5}$ & 0.0093 $\pm$ 0.0044 \\ 
       & 35.2 &  3225 &  905 & (4.5 $\pm$ 0.4) 10$^{-5}$ & 0.0084 $\pm$ 0.0036 \\ 
       & 35.2 &  4864 & 1139 & (4.2 $\pm$ 0.4) 10$^{-5}$ & 0.0077 $\pm$ 0.0026 \\ 
       & 35.2 & 10218 & 3289 & (3.9 $\pm$ 0.2) 10$^{-5}$ & 0.0064 $\pm$ 0.0025 \\ 
       & 35.2 & 34799 &23530 & (2.7 $\pm$ 0.1) 10$^{-5}$ & 0.0042 $\pm$ 0.0017 \\ [0.5ex]
 52499 & 2.9  &   127 &   35 & (4.7 $\pm$ 0.4) 10$^{-4}$ & 0.0823 $\pm$ 0.0247 \\ 
       & 2.9  &   182 &   37 & (3.5 $\pm$ 0.3) 10$^{-4}$ & 0.0721 $\pm$ 0.0153 \\ 
       & 2.9  &   369 &  103 & (3.4 $\pm$ 0.2) 10$^{-4}$ & 0.0584 $\pm$ 0.0105 \\ 
       & 2.9  &  1131 &  321 & (1.8 $\pm$ 0.1) 10$^{-4}$ & 0.0276 $\pm$ 0.0045 \\ 
       & 2.9  &  3833 & 1076 & (1.2 $\pm$ 0.5) 10$^{-4}$ & 0.0198 $\pm$ 0.0034 \\ 
       & 2.9  &  5610 & 1198 & (9.4 $\pm$ 0.6) 10$^{-5}$ & 0.0178 $\pm$ 0.0036 \\ 
       & 2.9  & 12893 & 4821 & (6.0 $\pm$ 0.4) 10$^{-5}$ & 0.0140 $\pm$ 0.0056 \\ 
       &  5.3 &   185 &   37 & (2.8 $\pm$ 0.4) 10$^{-4}$ & 0.0913 $\pm$ 0.0213 \\ 
       &  5.3 &   362 &  108 & (2.6 $\pm$ 0.2) 10$^{-4}$ & 0.0504 $\pm$ 0.0122 \\ 
       &  5.3 &   461 &   76 & (2.0 $\pm$ 0.2) 10$^{-4}$ & 0.0488 $\pm$ 0.0124 \\ 
       & 5.3  &  1082 &  297 & (1.5 $\pm$ 0.1) 10$^{-4}$ & 0.0394 $\pm$ 0.0055 \\ 
       & 11.0 &   520 &  132 & (1.6 $\pm$ 0.2) 10$^{-4}$ & 0.0349 $\pm$ 0.0101 \\ 
       & 11.0 &  1226 &  305 & (1.1 $\pm$ 0.1) 10$^{-4}$ & 0.0298 $\pm$ 0.0054 \\  
       & 11.0 &  4029 & 1095 & (6.8 $\pm$ 0.5) 10$^{-5}$ & 0.0225 $\pm$ 0.0031 \\  
       & 11.0 &  6037 & 1438 & (6.2 $\pm$ 0.4) 10$^{-5}$ & 0.0165 $\pm$ 0.0029 \\  
       & 11.0 & 13224 & 4429 & (5.4 $\pm$ 0.3) 10$^{-5}$ & 0.0128 $\pm$ 0.0025 \\ [0.5ex]
\tableline
\end{tabular}
}
\end{table}
\addtocounter{table}{-1}

\clearpage

\begin{table}
\caption{(continued)}  
\centerline{
\small
\begin{tabular}{@{}ccr@{\,$\pm$\,}lcc@{}}
\multicolumn{3}{c}{~} \\ [-2.ex]   
\tableline \tableline
\multicolumn{3}{c}{~} \\ [-2.5ex]                                                
MJD & Sky Level & \multicolumn{2}{c}{Signal Level} &
 \multicolumn{1}{c}{CTI} & Centroid Shift  \\
    & (e$^-$/pixel) & \multicolumn{2}{c}{(e$^-$)} & (pixel$^{-1}$) &
 (pixel) \\ [0.5ex] 
\tableline 
\multicolumn{3}{c}{~} \\ [-2.ex]   
 52499 & 32.6 &   483 &  111 & (1.1 $\pm$ 0.3) 10$^{-4}$ & 0.0310 $\pm$ 0.0126 \\  
       & 32.6 &   507 &  133 & (9.1 $\pm$ 0.2) 10$^{-5}$ & 0.0332 $\pm$ 0.0108 \\  
       & 32.6 &  1237 &  312 & (6.2 $\pm$ 0.8) 10$^{-5}$ & 0.0251 $\pm$ 0.0057 \\  
       & 32.6 &  4067 & 1108 & (5.2 $\pm$ 0.5) 10$^{-5}$ & 0.0239 $\pm$ 0.0033 \\  
       & 32.6 &  5986 & 1344 & (5.1 $\pm$ 0.4) 10$^{-5}$ & 0.0173 $\pm$ 0.0035 \\  
       & 32.6 & 13158 & 4413 & (4.6 $\pm$ 0.2) 10$^{-5}$ & 0.0145 $\pm$ 0.0022 \\  
       & 32.6 & 42713 & 8317 & (2.6 $\pm$ 0.2) 10$^{-5}$ & 0.0104 $\pm$ 0.0014 \\  [0.5ex]
 52885 &  3.4 &   141 &   27 & (4.0 $\pm$ 0.4) 10$^{-4}$ & 0.0979 $\pm$ 0.0308 \\ 
       &  3.4 &   250 &   55 & (3.5 $\pm$ 0.3) 10$^{-4}$ & 0.0863 $\pm$ 0.0152 \\ 
       &  3.4 &   372 &  107 & (2.9 $\pm$ 0.2) 10$^{-4}$ & 0.0709 $\pm$ 0.0125 \\ 
       &  3.4 &   468 &   75 & (2.3 $\pm$ 0.3) 10$^{-4}$ & 0.0643 $\pm$ 0.0153 \\ 
       &  3.4 &  1118 &  310 & (1.9 $\pm$ 0.1) 10$^{-4}$ & 0.0473 $\pm$ 0.0058 \\ 
       &  3.4 &  3862 & 1061 & (1.3 $\pm$ 0.1) 10$^{-4}$ & 0.0329 $\pm$ 0.0043 \\ 
       &  3.4 &  5630 & 1207 & (1.0 $\pm$ 0.1) 10$^{-4}$ & 0.0270 $\pm$ 0.0050 \\
       &  3.4 & 14176 & 4217 & (5.6 $\pm$ 0.5) 10$^{-5}$ & 0.0145 $\pm$ 0.0038 \\ 
       &  3.4 & 34584 &13298 & (4.0 $\pm$ 0.3) 10$^{-5}$ & 0.0085 $\pm$ 0.0031 \\ 
       & 11.3 &   517 &   74 & (2.0 $\pm$ 0.2) 10$^{-4}$ & 0.0442 $\pm$ 0.0130 \\
       & 11.3 &  1212 &  305 & (1.3 $\pm$ 0.1) 10$^{-4}$ & 0.0294 $\pm$ 0.0046 \\ 
       & 11.3 &  4077 & 1100 & (8.1 $\pm$ 0.5) 10$^{-5}$ & 0.0236 $\pm$ 0.0033 \\ 
       & 11.3 &  6155 & 1432 & (7.4 $\pm$ 0.4) 10$^{-5}$ & 0.0204 $\pm$ 0.0034 \\ 
       & 11.3 & 15001 & 4020 & (5.6 $\pm$ 0.2) 10$^{-5}$ & 0.0134 $\pm$ 0.0019 \\ 
       & 11.3 & 35832 &18767 & (3.3 $\pm$ 0.2) 10$^{-5}$ & 0.0076 $\pm$ 0.0015 \\ 
       & 15.5 &   540 &  137 & (1.4 $\pm$ 0.3) 10$^{-4}$ & 0.0337 $\pm$ 0.0153 \\ 
       & 15.5 &  1213 &  304 & (1.0 $\pm$ 0.1) 10$^{-4}$ & 0.0295 $\pm$ 0.0045 \\ 
       & 15.5 &  4064 & 1084 & (6.3 $\pm$ 0.5) 10$^{-5}$ & 0.0211 $\pm$ 0.0039 \\
       & 15.5 &  6086 & 1416 & (6.3 $\pm$ 0.4) 10$^{-5}$ & 0.0176 $\pm$ 0.0036 \\
       & 15.5 & 15225 & 4272 & (5.4 $\pm$ 0.3) 10$^{-5}$ & 0.0139 $\pm$ 0.0021 \\
       & 15.5 & 37163 &18658 & (3.1 $\pm$ 0.2) 10$^{-5}$ & 0.0082 $\pm$ 0.0022  \\ [0.5ex]
\tableline 
\end{tabular}
}
\end{table}

\clearpage

\begin{table}
\caption{Best-fit Values of Coefficients in Functional Form of Imaging
  CTI (Eq.~\ref{eq:imagCTIform}).} 
\label{t:imagcticoeff}
\centerline{
\small
\begin{tabular}{@{}cc@{~~~~~~~~}l@{}}
\multicolumn{3}{c}{~} \\ [-2.5ex]   
\tableline \tableline
\multicolumn{3}{c}{~} \\ [-2.2ex]                                                
Coefficient & Value & Description \\ [0.5ex] 
\tableline 
\multicolumn{3}{c}{~} \\ [-2.ex]   
$a$ & (1.33  $\pm$  0.03)\rlap{ 10$^{-4}$} & CTI normalization \\
$b$ & 0.54  $\pm$  0.01 & Count level dependence \\
$c$ & 0.205  $\pm$  0.006 & Time dependence of CTI \\
$d$ & 0.05  $\pm$  0.02 & Fraction of dependence on background (vs.\
 background/flux ratio) \\
$e$ & 0.82  $\pm$  0.05 & Multiplicative factor for background level dependence \\
$f$ & 3.60  $\pm$  0.07 & Multiplicative factor of background/flux count
 ratio dependence \\
$g$ & 0.21  $\pm$  0.02 & Power of background/flux count ratio \\ [0.5ex]
\tableline 
\end{tabular}
}
\end{table}

%% If you use the table environment, please indicate horizontal rules using
%% Do not put multiple tabular environments within a single table.
%% The optional \label should appear inside the \caption command.

%% Tables may also be prepared as separate files. See the accompanying
%% sample file table.tex for an example of an external table file.
%% To include an external file in your main document, use the \input
%% command. Uncomment the line below to include table.tex in this
%% sample file. (Note that you will need to comment out the \documentclass,
%% \begin{document}, and \end{document} commands from table.tex if you want
%% to include it in this document.)


%% The following command ends your manuscript. LaTeX will ignore any text


\begin{thebibliography}{}

\bibitem[Bohlin(1999)]{bohl99} 
Bohlin, R. C., 1999, Instrument Science Report STIS 1999-07 (Baltimore: STScI)
\bibitem[Bohlin et al.(1997)Bohlin, Collins, \& Gonnella]{bohl+97} 
Bohlin, R. C., Collins, N., \& Gonnella, A., 1997, Instrument Science Report
 STIS 1997-14 (Baltimore: STScI)
\bibitem[Bohlin \& Goudfrooij(2003)]{bohgou03} 
Bohlin, R. C., \& Goudfrooij, P., 2003, Instrument Science Report STIS 2003-03R 
 (Baltimore: STScI) 
\bibitem[Bristow \& Alexov(2002)]{brisalex02}
Bristow, P., \& Alexov, A., 2002, ST-ECF Instrument Science Report CE-STIS
 2002-001 (Garching: ST-ECF) 
\bibitem[Dolphin(2000)]{dolp00} 
Dolphin, A. E., 2000, \pasp, 112, 1397
\bibitem[Dressel et al.(2006)]{dres+06}
Dressel, L. L., Barrett, P. E., Goudfrooij, P., \& Hodge, P. E., 2006,
  in Proc.\ 2005 HST Calibration Workshop, eds.~A. M. Koekemoer, P. Goudfrooij, \&
  L. L. Dressel (Baltimore: STScI), 267
\bibitem[Gilliland et al.(1999)Gilliland, Goudfrooij, \& Kimble]{gill+99} 
Gilliland, R. L., Goudfrooij, P., \& Kimble, R. A., 1999, \pasp, 111, 1009 
\bibitem[Goudfrooij \& Walsh(1997)]{gouwal97} 
Goudfrooij, P., \& Walsh, J. R., 1997, Instrument Science Report STIS 1997-09 
 (Baltimore: STScI) 
\bibitem[Goudfrooij \& Kimble(2003)]{goukim03} 
Goudfrooij, P., \& Kimble, R. A., 2003, in 2002 HST Calibration Workshop, eds.\
 S. Arribas, A. Koekemoer, \& B. C. Whitmore (Baltimore: STScI), p.\ 105
%\bibitem[Goudfrooij et al.(2006)]{goud+06a} 
%Goudfrooij, P., Ma\'{\i}z Apell\'aniz, J., Brown, T. M., \& Kimble, R. A.,
% 2006, Instrument Science Report STIS 2006-01 (Baltimore: STScI) 
\bibitem[Goudfrooij \& Bohlin(2006)]{gouboh06} 
Goudfrooij, P., \& Bohlin, R. C., 2006, Instrument Science
 Report STIS 2006-03 (Baltimore: STScI) 
%\bibitem[Goudfrooij et al.(2006)Goudfrooij, Bohlin, \& Ma\'{\i}z
%  Apell\'aniz]{goud+06a} 
%Goudfrooij, P., Bohlin, R. C., \& Ma\'{\i}z Apell\'aniz, J., 2006, in
% 2005 HST Calibration Workshop, eds.\ A. Koekemoer, P. Goudfrooij, \&
% L. L. Dressel (Baltimore: STScI), p.\ 289 
\bibitem[Hardy et al.(1998)Hardy, Murowinski, \& Deen]{hard+98} 
Hardy, T., Murowinski, R., \& Deen, M. J., 1998, IEEE Trans.\ 
 Nucl. Sci., 45, 154
\bibitem[Janesick et al.(1991)]{jane+91} 
Janesick, J., Soli, G., Elliot, T., \& Collins, S., 1991, SPIE Electronic
 Imaging and Technology Conference on ``Solid State Optical Sensors II,'' 1147 
\bibitem[Kimble et al.(1994)]{kimb+94} 
Kimble, R. A., Brown, L., Fowler, W. B., Woodgate, B. E., Yagelowich, J. J.,
 et al., 1994, Proc.\ SPIE, 2282, p.\ 169
\bibitem[Kimble et al.(2000)Kimble, Goudfrooij, \& Gilliland]{kimb+00} 
Kimble, R. A., Goudfrooij, P., \& Gilliland, R. L., 2000, Proc.\ SPIE, 4013, 
 p.\ 532
\bibitem[Kim Quijano et al.(2004)]{kimq+04} 
Kim Quijano, J., et al., 2004, STIS Instrument Handbook (Baltimore:
 STScI) 
\bibitem[Koslowski et al.(2006)]{kosl+06}
Koslowski, S., Wo\'zniak, P. R., Mao, S., Smith, M. C., Sumi, T., Vestrand,
 W. T., \& Wyrzykowski, L., 2006, MNRAS, doi:10.1111/j,1365-2966.2006.10487.x
\bibitem[Leitherer \& Bohlin(1997)]{leiboh97} 
Leitherer, C., \& Bohlin, R. C., 1997, Instrument Science Report STIS 97-13 
 (Baltimore: STScI)
\bibitem[Malumuth(1996)]{malu96} 
Malumuth, E. M., 1996, Pre-launch STIS Analysis Report \#55
 (Greenbelt: NASA/GSFC)
\bibitem[McGrath et al.(1999)McGrath, Busko, \& Hodge]{mcgr+99} 
McGrath, M. A., Busko, I., \& Hodge, P., 1999, Instrument Science Report 
 STIS 99-03 (Baltimore: STScI)
\bibitem[McNamara et al.(2003)McNamara, Harrison, \& Anderson]{mcna+03}
McNamara, B. J., Harrison, T. E., \& Anderson, J., 2003, \apj, 595, 187
\bibitem[Piatek et al.(2005)]{piat+05}
Piatek, S., Carlton, P., Bristow, P., Olszewski, E. W., Harris, H. C.,
 Mateo, M., Minniti, D., \& Tinney, C. G., 2005, \aj, 130, 95
\bibitem[Proffitt(2006)]{prof06} 
Proffitt, C. R.,  2006, in Proc. 2005 HST Calibration Workshop,
  eds.~A. M. Koekemoer, P. Goudfrooij, \& L. L. Dressel (Baltimore: STScI), 199
\bibitem[Riess et al.(1999)Riess, Biretta, \& Casertano]{ries+99} 
Riess, A., \& Mack, J., 2004, Instrument Science Report ACS 2004-06
 (Baltimore: STScI) 
\bibitem[Stetson(1987)]{stet87} 
Stetson, P. B., 1987, \pasp, 99, 191 
\bibitem[Stys et al.(2004)Stys, Bohlin, \& Goudfrooij]{stys+04} 
Stys, D. J., Bohlin, R. C., \& Goudfrooij, P., 2004, Instrument Science Report 
 STIS 2004-04 (Baltimore: STScI)
\bibitem[Whitmore et al.(1999)Whitmore, Heyer, \& Casertano]{whit+99} 
Whitmore, B. C., Heyer, I., \& Casertano, S., 1999, \pasp, 111, 1559
\bibitem[Woodgate et al.(1998)]{wood+98} 
Woodgate, B. E., Kimble, R. A., Bowers, C. W., Kraemer, S., Kaiser, M. E., 
 et al., 1998, \pasp, 110, 1183

\end{thebibliography}
\end{document}